\documentclass[useAMS,referee,usenatbib]{mn2e}

\usepackage{graphicx}

\def\ms1008{$\mathrm{MS\,1008}$}
\def\re{${r_{\rm e}}$}

\def\lre{${\log r_{\rm e}}$}
\def\mie{${< \! \mu \! >_{\rm e}}$} 
\def\mi0{${\mu_0}$} 
\def\sn{$n$}
\def\lsn{$\log n$}
\def\KS{$K(S)$}
\def\KI{$K(I)$}
\def\bls{$\rm CBLS$}
\def\amls{$\rm CAMLS$}
\def\orls{$\rm CORLS$}
\def\gmls{$\rm CGMLS$}
\def\cls{$\rm CLS$}
\def\sln{$\sigma_{\log n}$}
\def\sir{$\sigma^{\rm i}_{\log r_{\rm e}}$}

\title[Photometric Planes at $z \! \sim \! 0.3$]{New insights into the
   structure of early-type galaxies:  the Photometric Plane at $z \sim
   0.3$\footnote{Based on observations  collected at European Southern
   Observatory (ESO ID.   60.A9203, 60.A-9021, 60.O-9025, 66.A-0316).}
   }       \author[F.      La      Barbera     et      al.]{F.      La
   Barbera$^{1}$\thanks{E-mail:        labarber@na.astro.it},       G.
   Covone$^{2}$\footnotemark[1]\thanks{E-mail:
   giovanni.covone@oamp.fr},       G.        Busarello$^{1}$,       M.
   Capaccioli$^{1,3}$,     C.P.     Haines$^{1}$,     \newauthor    A.
   Mercurio$^{1}$,  P.  Merluzzi$^{1}$  \\  $^{1}$INAF -  Osservatorio
   Astronomico  di  Capodimonte,  Via  Moiariello 16,  Napoli,  80124,
   Italy.\\ $^{2}$CNRS -  Laboratoire d'Astrophysique de Marseille, BP
   8,  Traverse  du Siphon,  13376  Marseille  Cedex  12, France.   \\
   $^{3}$Physics Department, Universit\`a Federico II, Napoli, Italy.}
\begin{document}

\date{ }

\pagerange{\pageref{firstpage}--\pageref{lastpage}} \pubyear{2005}

\maketitle

\label{firstpage}

\begin{abstract}
We study the Photometric Plane  (PHP), namely the relation between the
effective radius  \re, the mean surface brightness  within that radius
\mie,  and  the  Sersic  index  \sn,  in optical  ($R$  and  $I$)  and
near-infrared ($K$)  bands for a  large sample of  early-type galaxies
(ETGs)  in the  rich  cluster \ms1008  \,  at $z  =  0.306$.  The  PHP
relation is  $\log r_{\rm e} \!  =  \!  (1.07 \!  \pm  \!  0.06) \cdot
\log n \! + \!  (0.219 \!   \pm \!  0.009) \cdot < \!  \mu \!>_{\rm e}
+ \, const$, with an intrinsic dispersion of $ \sim 32 \%$ in \re, and
turns out  to be independent  of waveband.  This result  is consistent
with the fact  that internal colour gradients of ETGs  can have only a
mild dependence on galaxy luminosity (mass).  There is no evidence for
a  significant  curvature  in the  PHP.   We  show  that this  can  be
explained if this relation origins  from a systematic variation of the
specific entropy of  ETGs along the galaxy sequence,  as was suggested
from  previous  works~\citep{MLC01}.   Indeed, considering  spherical,
non-rotating, one-component  galaxy models, we find  that the specific
entropy is  exactly a  linear combination of  \lre, \mie \,  and \lsn.
The intrinsic  scatter of  the PHP is  significantly smaller  than for
other purely photometric relations,  such as the Kormendy relation and
the  photometric  Fundamental Plane,  which  is  constructed by  using
colours in place of velocity dispersions.  The scatter does not depend
on the  waveband and  the residuals about  the plane do  not correlate
with residuals of the  colour-magnitude relation.  This implies either
that the  scatter of the PHP  does not origin  from stellar population
parameters or that it is due  to a combined effect of such parameters.
Finally, we compare  the coefficients of the PHP at  $z \sim 0.3$ with
those of ETGs at  $z \sim 0$, showing that the PHP  is a valuable tool
to  constrain the  luminosity evolution  of ETGs  with  redshift.  The
slopes of the PHP do not change significantly with redshift, while the
zero-point  is consistent  with  cosmological dimming  of the  surface
brightness in an expanding universe  plus the passive fading of galaxy
stellar populations with  a high formation redshift ($z_{\rm  f} \!  >
\!  1$--$2$).
\end{abstract}


\section{Introduction}
\label{intr}
Global properties of early-type galaxies (ETGs), such as luminosities,
colours,  radii, line  indices and  velocity dispersions,  are tightly
correlated,   implying   that    stellar   population   as   well   as
dynamical/structural properties  of these  systems vary smoothly  as a
function  of  their  mass.   One  of  the  most  well  known  of  such
correlations is the Fundamental Plane (FP), which is usually expressed
as  a relation  among  the  effective parameters  of  ETGs, i.e.   the
effective (half-light)  radius \re \, and the  mean surface brightness
within  that   radius  \mie,  and  the   central  velocity  dispersion
$\sigma_0$~\citep{DjD87, DLB87}.   The main characteristics  of the FP
are its small intrinsic dispersion, in the range of $0.06$--$0.13$~dex
($14-30 \!   \%$) in  $ r_{\rm e}$  \, \citep{JFK96, PDdC98},  and its
tilt, i.e.   the deviation  of its slopes  from those predicted  for a
virialized  family of homologous  systems with  constant mass-to-light
ratios~\citep{BUS97}.  Several works have studied the FP to $z \sim 1$
(e.g.  \citealt{TRAN04, WUY04}  and references therein).  However, due
to the high  demand of observing time for  the measurement of velocity
dispersions, these  studies have been  based only on small  samples of
galaxies.   For  this reason,  different  efforts  have  been made  to
construct  correlations among purely  photometric parameters  of ETGs,
such  as  the  mean   surface  brightness  (or  the  luminosity)--size
relation, also  known as the Kormendy relation  \citep{KOR77}, and the
size--profile  shape  relation \citep{CCD93}.   The  main drawback  of
these relations  is that  their intrinsic dispersion  is significantly
larger  with respect  to  that of  the  FP.  Since  biases on  fitting
coefficients become larger as  the dispersion of observed correlations
increases,  selection effects  are a  crucial issue  for the  study of
purely photometric correlations of ETGs.

Another interesting correlation among global properties of ETGs is the
so-called  photometric  plane  (hereafter  PHP,  see  \citealt{GRA02},
GRA02, and references therein),  that is the correlation among radius,
surface brightness and Sersic index  (shape parameter) \sn \, of ETGs.
As shown  by GRA02, the PHP  could have an intrinsic  scatter which is
comparable or  slightly larger than  that of the FP,  therefore making
this relation an  interesting tool to measure galaxy  distances and to
analyze the  properties of galaxies at different  redshifts.  To date,
however,  a detailed  analysis  of the  use  of the  PHP for  studying
samples  of  distant galaxies  has  not  been done.   \citet[hereafter
LMB04]{LMB04}  firstly attempted  to  derive  the PHP  for  ETGs in  a
cluster at  $z \sim 0.2$, finding  that a PHP relation  seems to exist
also at this  redshift.  The other few existing works  on the PHP have
only analyzed samples of galaxies at $z \sim 0$.  Moreover, a straight
comparison of results  of these works cannot be  done, since they have
derived the PHP in different wavebands, by using structural parameters
defined  in different ways,  and for  differently selected  samples of
galaxies.  \citet{KWK00,  KRP04} derived the $K$-band PHP  for ETGs in
the  Coma cluster  and  in  nearby groups  using  the central  surface
brightness, $\mu_0$, the logarithm  of the effective radius, \lre, and
the  logarithm   of  the   Sersic  index,  \lsn.    \citet{LGM99}  and
\citet{MLC00}  (hereafter LGM99  and MLC00  respectively)  studied the
optical  PHP of  ETGs in  clusters at  $z \sim  0$ using  $\mu_0$, the
inverse of the Sersic index, $\nu = 1/n$, and the scale--length of the
Sersic law (see eq.~7 of LNG00).  All these studies found that the PHP
is actually a surface, ETGs  populating a curved manifold in the space
of structural parameters.  On the other hand, GRA02 derived the PHP in
the $B$  band for Virgo  and Fornax ETGs,  finding that ETGs  follow a
linear relation  among \lre, \mie \,  and \lsn.  As  noticed by GRA02,
since the  Sersic law fit may  not always provide  realistic values of
the central surface brightness,  especially for galaxies with high \sn
\, values, the  use of \mie \, should generally  be preferred to \mi0.
We  note  that  this  problem  becomes even  more  important  at  high
redshift, where, due to seeing effects, the measurement of \mi0 \, can
require a large extrapolation of the light profile.

A  possible  explanation for  the  existence  of  a correlation  among
structural  parameters of  ETGs  has been  suggested  by LGM99,  MLC00
and~\citet[hereafter  MLC01]{MLC01}.   Using spherical,  non-rotating,
one-component models of ETGs, these works argued that the PHP relation
origins from  a relation between the  mass of ETGs  and their specific
entropy.  MLC00  showed that  the entropy-mass relation  is consistent
with  that expected in  a dissipation--less  merging scheme  of galaxy
evolution, where merging produces a higher level of disorder (entropy)
in larger galaxies.  This interpretation is supported by the fact that
the PHPs  of dwarf  and normal  ellipticals appear to  be offset  in a
direction  of larger entropy  for normal  galaxies~\citep{KRP04}.  The
connection  between the  PHP and  processes such  as dissipation--less
merging makes  it even  more interesting to  analyze this  relation at
different  redshifts.  If  ETGs were  mostly assembled  at $z  \sog 1$
(e.g.~\citealt{KAU95}), one  would expect  that at lower  redshift the
slopes  of the PHP  do not  change significantly  with $z$,  while its
zero-point varies  accordingly to  the fading of  stellar populations.
Just as  for the FP, this would  imply that the zero-point  of the PHP
could be used to measure the formation epoch of stellar populations in
ETGs  and,  perhaps,  cosmological  parameters.   Since  at  different
redshifts  structural parameters  are derived  in  different restframe
bands, the evolution with $z$ of  the PHP can only be addressed on the
basis of  an accurate knowledge  of the wavelength dependence  of this
relation.  On  the other hand, the  dependence of the  PHP on waveband
carries interesting information by  itself, depending on how the ratio
of structural parameters between  different wavebands varies along the
galaxy  sequence.   This   variation  indicates  how  internal  colour
gradients of ETGs  depend on mass, which is  an important discriminant
of galaxy formation scenarios~\citep{PDI90}.

In the present  paper, we derive the PHP in  optical and NIR wavebands
for a  large sample of ETGs  belonging to the cluster  of galaxies ClG
1008-1224, also known as MS1008-1224 (hereafter \ms1008), a rich X-ray
bright cluster  at $z  = 0.306$~\citep{LEW99}, originally  detected in
the  Einstein  Medium   Sensitivity  Survey~\citep{GiL94}.   For  this
cluster, a  unique wealth of  multi-wavelength data is  available from
the ESO archive~\footnote{http://archive.eso.org}, including very deep
$UBVRIJHK$  photometry taken  with VLT  FORS  and ISAAC  and NTT  SOFI
instruments  in excellent seeing  conditions.  The  data used  for the
present  analysis consist  of  the  R-, I-  and  K-band photometry  of
\ms1008.  This unique data-set allows us to perform for the first time
a  homogeneous and accurate  multi-wavelength analysis  of the  PHP at
intermediate redshifts.
In  the present  paper we  study (i)  the characteristics  of  the PHP
relation (i.e.  its coefficients, scatter  and shape) at $z \sim 0.3$,
accounting for selection effects as well as for the fitting procedure;
(ii) the  waveband dependence of the  PHP; and (iii)  the variation of
the PHP coefficients from $z \sim 0.3$ to $z \sim 0$.  We consider the
following representation of the PHP:
\begin{equation}
{ \log r_{\rm e} = a \log n + b < \! \mu \! >_{\rm e} +c },
\label{PHPEQ}
\end{equation}
where $a$  and $b$  are the slopes  and $c$  is the zero-point  of the
relation.

The layout of  the paper is as follows.  In  Section~2 we describe the
samples  used  in  the   analysis,  while  in  Section~3  the  surface
photometry is presented.  Section~4 deals with the problems related to
the fit of the PHP.  In Section~5 we derive the PHP at $z \sim 0.3$ in
both optical and NIR wavebands.   In Section~6 we present the waveband
and  redshift  dependence  of  the  PHP.  The  variation  of  the  PHP
coefficients with  redshift is analyzed trough  a comparative analysis
of the PHP  at $z \sim 0.3$ with  that of ETGs in clusters  at $z \sim
0$.   The main  results are  then discussed  in Section~7.   A summary
follows in Section~8.   In the present paper we  assume a $\Lambda$CDM
cosmology  with  $H_0=  {\rm  70~Km~s^{-1}~Mpc^{-1}  }$,  $\Omega_{\rm
m}=0.3$ and $\Omega_\Lambda=0.7$).

\section{The samples}
\label{SAMPLES}
The galaxies used for the  present study belong to the cluster \ms1008
\, at $z = 0.306$ and were selected on the basis of a large wavelength
baseline,  including $UBVRIJHK$ photometry  for a  field of  $\sim 6.8
\arcmin \times  6.8 \arcmin$  around the cluster  center ($\alpha_{\rm
J2000}$=10:10:34.1,           $\delta_{\rm          J2000}$=-12:39:48,
see~\citealt{GMS90}).  This dataset, which  was retrieved from the ESO
archive  and  reduced  by  the  authors  as  described  in  Covone  et
al.~(2005,  in preparation,  hereafter  CLB05), was  used to  estimate
photometric  redshifts  for all  the  sources  in  the cluster  field.
Details on photometric redshifts can  be found in CLB05, while we give
here  only  the  relevant  information.   In  CLB05  we  compared  the
photometric redshifts with spectroscopic redshifts available for $N \!
\sim  \!    70$  galaxies   in  the  cluster   field  from   the  CNOC
survey~\citep{YEM98}.   The mean  offset  and the  rms of  differences
between  spectroscopic  and photometric  redshifts  turned  out to  be
$0.004 \pm 0.005$  and $\delta \!  z \!   \sim \!  0.04$ respectively.
The quantity $\delta \!  z$  gives an estimate of the typical accuracy
on photometric redshifts.  We  selected as cluster members the objects
with  photometric redshifts in  a range  of $\pm  0.08$ (i.e.,  $\pm 2
\delta \!  z $) around the cluster redshift.

\begin{figure*}
\centering
\includegraphics[angle=0,width=15cm,height=15cm]{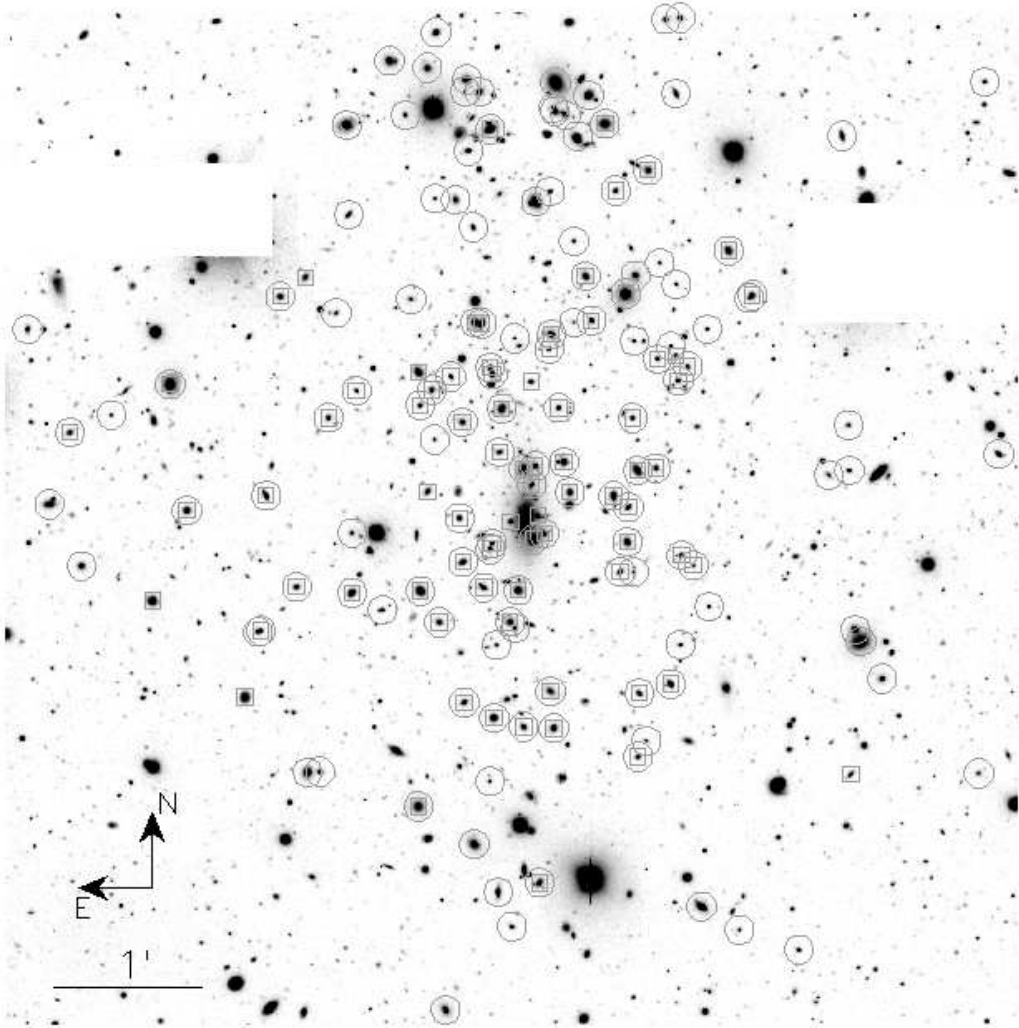}
\caption{FORS1 R-band image  of \ms1008.  The image covers  an area of
$\sim   6.8   \arcmin  \times   6.8   \arcmin$   around  the   cluster
center. Galaxies  included in the optical  ($R$ and $I$  bands) and in
the K-band samples are marked by circles and squares respectively.  }
\label{SAM}
\end{figure*}

For  the  present study,  we  use only  the  $R$-,  $I$- and  $K-$band
photometry  of \ms1008,  whose depth  and resolution  are  suitable to
obtain structural parameters for a  fair sample of galaxies.  The $R$-
and  $I$-band  data were  taken  with  VLT--FORS2  at ESO,  while  the
$K$-band  imaging,  including ESO--VLT  and  ESO--NTT photometry,  was
taken with ISAAC and SOFI instruments respectively.  The $R$- and $I$-
band images cover  a field of $6.8 \arcmin \times  6.8 \arcmin$ with a
pixel scale of $0.201$~$\rm  \arcsec /pixel$, while the $K$-band SOFI,
hereafter  \KS, and $K$-band  ISAAC, hereafter  \KI, images  cover two
fields of $5 \arcmin \times 5 \arcmin$ and $\rm 2.5 \arcmin \times 2.5
\arcmin$ around  the cluster center with pixel  scales of $0.288$~$\rm
\arcsec /pixel$ and $  0.147$~$\rm \arcsec / pixel$ respectively.  The
exposure times  of the  $R$-, $I$-, \KS-  and \KI- images  were $1.5$,
$1.125$,  $5.7$ and $1.4$~$\rm  hours$, while  mean seeing  sizes were
$0.8  \arcsec$,  $0.9  \arcsec$,   $0.8  \arcsec$  and  $0.5  \arcsec$
respectively.   We  derive  the  PHP  for four  different  samples  of
galaxies, in  the $R$-, $I$-,  \KS- and \KI-band  images respectively.
Galaxies in each sample were  selected as follows.  In order to obtain
reliable structural  parameters, we selected galaxies down  to a given
magnitude cut, which was established by deriving structural parameters
from  simulated  galaxy  images  (see sec.3.3  of~\citealt{LBM02}  for
details).   For each  waveband, we  estimated the  magnitude  limit to
which systematic  uncertainties on  effective radius and  Sersic index
are  expected to  be negligible.   Hence, we  considered  only cluster
members brighter  than $R=22.0$, $I=21.1$,  and $K=17.8$ in  the $R$-,
$I$-,  and  $K$-band  images  respectively.  We  also  excluded  those
objects  very close  to bright  saturated  stars in  the field,  since
structural parameters  of those  galaxies would have  large systematic
uncertainties  due  to  background  subtraction problems.   A  further
selection was  applied {\it  a posteriori}, by  selecting as  ETGs the
objects with Sersic  index $n \!  > \!  2$  and excluding objects with
uncertainties  on structural  parameters greater  than $100  \%$.  The
adopted Sersic index cut corresponds to exclude disk-dominated systems
from the present analysis (see e.g.~\citealt{vDF98}).  The fraction of
excluded objects  was smaller  than $7 \%$  for each band.   The final
$R$, $I$, \KS \, and \KI \, samples include $N=129$, $ N=123$, $ N=68$
and $ N=50$  galaxies respectively, with $N=112$ objects  in common to
the $R-$ and $I-$band images, and $N=38$ galaxies are in common to the
\KS \, and \KI samples.  The position of these galaxies in the cluster
field is shown in Fig.~\ref{SAM}.

We point  out that the  above selections allow  us to obtain  for each
sample two sharp selection cuts in the space of structural parameters,
i.e.  the magnitude  limit and the Sersic index  cut.  This is crucial
for an  accurate correction of PHP coefficients  for selection effects
(Section~\ref{SEL}).
  
\section{Derivation of Structural parameters}
\label{SPAR}
Structural  parameters were derived  for galaxies  in the  $R$-, $I$-,
\KS- \, and \KI-band images of \ms1008 \, by using the two-dimensional
fitting  method  (see \citealt{LBM02}  and  references therein).   The
surface  brightness  of  galaxies   was  modeled  by  the  Sersic  law
\citep{SER68}:
\begin{equation}
\mu(r) = \mu_0 +2.5 b_n \log \mathrm{e} \cdot (r / r_{\rm e})^{1/n}, 
\label{SLAW}
\end{equation}
where $b_n$ is  a constant, defined in such a way  that $r_{\rm e}$ is
the effective (half-light)  radius of the galaxy, r  is the equivalent
radius, $n$  is the Sersic  index (shape parameter), and  ${\mu_0}$ is
the central  surface brightness.  The constant $b_n$  can be estimated
by a power--law  in \sn~\citep{CIO99} with $b_n \sim  2n-1/3$ in first
approximation~\citep{CAP89}, or,  with higher accuracy ($<  1 \%$), by
the function  $b_n \sim  \exp[ (0.6950 +  \ln n)  - 0.1789/n ]  $ (see
LGM99).  The mean surface brightness within \re, \mie, is given by the
following formula (e.g.~\citealt{CIO99}):
\begin{equation}
\begin{array}{lll}

m_{\rm T} & = & -2.5 \log(2 \pi) + \mu_0 -5 \log(r_{\rm e}) + \\ & & + 5n \log
    b_n  -2.5  \log \Gamma(2n)  \\  &  = &    -2.5  \log(2 \pi)  -5
    \log(r_{\rm e}) + < \! \mu \! >_{\rm e}, \\
\label{magn}
\end{array}
\end{equation}
where $m_{\rm T}$  is the total magnitude of the  galaxy and $ \Gamma$
is the complete gamma function.

Galaxy  images  were  fitted  with  seeing--convolved  Sersic  models,
minimizing the function:
\begin{equation}
\mathrm{
\chi^2  = \sum_{i,j}  \left[ G_{i,j}  -  ( bg  + S  \otimes PSF_{i,j}  )
\right]^2, }
\label{CHI2}
\end{equation}
where  the  symbol $\otimes$  denotes  convolution,  $G_{i,j}$ is  the
galaxy surface brightness at pixel  $\left[ i,j \right]$, $ bg$ is the
local background  value, while $S$ and  $ PSF$ are the  Sersic and the
PSF models  respectively.  Seven output parameters  were provided from
the  fitting process:  the center  coordinates, the  effective radius,
\re, the  mean surface  brightness, \mie, the  Sersic index,  \sn, the
position angle, $ PA$, of major  axis and the axis ratio, $ b/a$.  For
each  galaxy, neighbor  objects  were masked  automatically using  the
ellipticity, position angle and isoarea parameters provided by running
SExtractor~\citep{BeA96}  on the corresponding  images.  In  the cases
where  SExtractor  failed  to  provide  reliable  estimates  of  these
parameters, such as  in crowded regions and/or in  the neighborhood of
extended  sources, masking  was performed  interactively.   Very close
galaxies were fitted simultaneously.

The PSFs of the  $R$, $I$, \KS \, and \KI \,  images were modeled by a
sum of 2D Moffat functions,  taking into account deviations of stellar
isophotes from circular symmetry. Details on PSF modeling are given in
Appendix~\ref{PSF}.  Uncertainties  on \lre, \mie \, and  \lsn \, were
estimated by  using numerical simulations and  by comparing structural
parameters  among the different  wavebands.  Details  can be  found in
Appendix~\ref{UNCER}.  Structural  parameters\footnote{Table~1 is only
available   in    electronic   form   or   via    anonymous   ftp   to
cdsarc.u-strasbg.fr at the  CDS.}  for galaxies in the  $R$-, $I$- and
$K$-band samples are reported in Table~1.  Since structural parameters
turned  out to  be fully  consistent  between the  \KS \,  and \KI  \,
samples (see Appendix~\ref{UNCER}), in  Table~1 we report the averages
of the \KS \, and \KI \, structural parameters, which were computed by
weighting  each value  with the  inverse square  of  the corresponding
uncertainty.

\begin{table*}
\setcounter{table}{1}
\large
\caption{Acronyms used for fitting methods. Since the fit coefficients
were corrected  for selection  effects (see Section~\ref{SEL}),  a prefix
{\it C} was used for each acronym.  }
\label{FITNAME}
\centering
\begin{tabular}{|l|l|}
\hline 
 CLS$_{\log r_{\rm e}}$ & Corrected weighted least square fit with
 dependent variable \lre.  \\ 
 CLS$_{< \! \mu \! >_{\rm e}}$ & Corrected weighted least square fit with
 dependent variable \mie \\ 
 CLS$_{\log n}$ & Corrected weighted least square fit with
 dependent variable \lsn \\ 
 $\rm CORLS$ & Corrected orthogonal weighted least-square fit \\ 
 $\rm CBLS$ & Corrected bisector least square fit \\ 
 $\rm CAMLS$ & Corrected arithmetic mean of the least-square coefficients \\ 
 $\rm CGMLS$ & Corrected geometric mean of the least-square coefficients \\ 
\hline
\end{tabular}
\end{table*}

\section{Fitting the Photometric Plane}
\label{FIT}
We derived the coefficients of the PHP by accounting for (i) selection
effects, i.e.   the cuts in Sersic  index and magnitude;  and (ii) the
correlated  uncertainties on  structural parameters.   Since different
fitting  methods  are  not  equally  affected by  these  effects  (see
\citealt{LBM00}, LBC00),  the PHP coefficients were  obtained for each
sample   by  using   different  fitting   methods,  as   described  in
Section~\ref{FITS}.  Correction for selection  effects was performed by a
Monte Carlo technique as detailed in Section~\ref{SEL}.

\subsection{Regression methods}
\label{FITS}
The  PHP coefficients were  derived by  using seven  different fitting
procedures:  three weighted  ordinary least-square  fits,  adopting as
dependent variable one  of the quantities \lre, \mie  \, and \lsn; the
orthogonal weighted least square fit, where the root mean square (rms)
of residuals  perpendicular to the  plane are minimized;  the bisector
least square (BLS)  fit (see LBC00); the arithmetic  and the geometric
means of  the ordinary  least-square coefficients.  In  the following,
acronyms will be  used for different fitting methods  as summarized in
Table~\ref{FITNAME}.

In each fitting procedure, we derive the coefficients $a$, $b$ and $c$
of  the  plane  (see  Eq.~\ref{PHPEQ}) and  its  intrinsic  dispersion
$\sigma^{\rm i}$.  For the  ordinary and orthogonal least square fits,
these quantities were derived by minimizing the following expression:
\begin{equation}
\begin{array}{rl}
\chi^2 = & - \ln L =\\
= & \sum_k \frac{r_k(a, b,c)^2}{\epsilon_k(a,b,c)^2+(\sigma^{i})^2}
+ 1/2 \ln \left[ \epsilon_k(a,b,c)^2+ (\sigma^{\rm i})^2  \right]
,
\label{CHIEQ}
\end{array}
\end{equation}
where $L$  is the likelihood  function, $r_k$ are the  residuals about
the  plane,   $\epsilon_k$  are   the  uncertainties  on   $r_k$,  and
$\sigma^{\rm  i}$ is  the intrinsic  dispersion of  the PHP  along the
direction, $\bmath  d_{\rm min}$, where residuals  are minimized.  The
terms $\epsilon_k$ are the uncertainties on $r_k$ and were obtained by
projecting the  covariance matrix of measurement  errors on structural
parameters  along $\bmath d_{\rm  min}$.  We  point out  that although
this procedure allows each point to be weighted with the corresponding
uncertainties,  it does  not  correct exactly  for  biases on  fitting
procedures which are due  to the correlation among such uncertainties.
It  is always  possible, in  fact, to  vary the  covariance  matrix of
measurement errors  on \lre,  \mie \, and  \lsn, without  changing its
projection on  a given  direction $\bmath d_{\rm  min}$.  In  order to
estimate the  amount of bias  which is due  to the correlation  of the
uncertainties,  we  adopted  the  MIST  fits  (LBC00)  which  provides
unbiased  values of  ordinary least-square  coefficients by  using the
mean covariance  matrix of uncertainties on the  three variables.  The
bias turned out to be always  smaller than $5 \%$ for each sample, and
therefore we did  not apply corrections for this  effect.  In order to
derive the quantities $a$, $b$, $c$ and $\sigma^{\rm i}$, we minimized
Eq.~\ref{CHIEQ}  by using a  Levenberg-Marquardt algorithm.   For what
concerns the  \bls, \amls \, and  \gmls \, fits,  the PHP coefficients
were  calculated from  those  of the  CLS  fits.  For  this reason,  a
different  weight  is given  to  each  galaxy  also in  these  fitting
methods.   Uncertainties on $a$,  $b$, $c$  and $\sigma^{\rm  i}$ were
estimated  by  the   bootstrap  method,  applying  $N=2500$  bootstrap
iterations.

\subsection{Selection effects}
\label{SEL}
Selection effects  were corrected for by  using numerical simulations,
producing  distributions   of  points  in  the   space  of  structural
parameters  which resembled those  of real  galaxies.  Points  in each
simulation  were  generated  as  follows.   Magnitudes  were  assigned
according to the luminosity function, which was modeled as a Schechter
function with slope  $\alpha$ and characteristic magnitudes $M^\star$.
The values of $\alpha$ and $M^\star$ were drawn from \citet{BML02} for
the  $R$  and $I$  bands,  and from  \citet{dP99}  for  the $K$  band.
Effective radii and mean  surface brightnesses were obtained according
to the luminosity-size relation,  whose slopes, zero-point and scatter
were derived from  the data of \ms1008.  Sersic  indices were assigned
using the PHP coefficients, $a$, $b$ and $c$, and its \lsn \, scatter,
\sln.   The values of  $a$, $b$,  $c$ and  \sln \,  were chosen  by an
iterative procedure.  For each  iteration, a simulated distribution in
the  space of structural  parameters was  constructed by  imposing the
same selection effects, i.e.  the  magnitude and Sersic index cuts, of
the real samples, and the  PHP coefficients were computed by using all
the fitting methods described  in Section~\ref{FITS}.  The values of $a$,
$b$, $c$  and \sln were  modified until the coefficients  derived from
the simulated samples matched those  of the real samples for $\it all$
the fitting methods. This procedure allowed us to achieve an excellent
match, with  an accuracy better than  $\sim 5 \%$ for  all the fitting
coefficients.   Once the  values of  $a$, $b$,  $c$ and  \sln  \, were
chosen, we  estimated the relative  variations of PHP  coefficients of
all the regression methods after  selection cuts were removed from the
simulated samples.  The variations were used as correction factors for
the   PHP    coefficients   of    \ms1008.    As   an    example,   in
Table~\ref{CORPHPR} we report the  correction factors for the $R$-band
sample of \ms1008.
We  see that  the  bias can  be  as large  as $\sim  -35  \%$ for  the
coefficient  $a$ of  the  CLS$_{\log  n}$ fit,  and  that it  strongly
depends  on the fitting  method, being  smaller than  $\sim 10  \%$ in
absolute  value  for  the \orls,  \bls  \,  and  \amls \,  fits.   The
uncertainties on the correction factors were estimated by changing the
input  parameters of  the  simulation algorithm  (e.g.   the value  of
$\alpha$)  according   to  their  uncertainties,   and  repeating  the
iteration procedure.  We  found that the bias on  the PHP coefficients
varies at most by  $\sim 4 \%$ for the coefficient $a$  in the \cls \,
fits.  This  is smaller with  respect to the typical  uncertainties on
PHP coefficients, and was therefore neglected.
\begin{table}
\large
\caption{Bias  on  the  $R$-band  PHP coefficients  due  to  selection
effects.}
\label{CORPHPR}
\centering
\begin{tabular}{|l|c|c|c|c|c|}
\hline
 & $\frac{\delta a}{a}$ & $ \frac{\delta b}{b}$ & $ \frac{\delta c}{c}$ & $\frac{\delta \sigma_{\log r_{\rm e}} }{ \sigma_{\log r_{\rm e}}} $ & $ \frac{\delta \sigma_{\log r_{\rm e}}^{\rm i} }{ \sigma_{\log r_{\rm e}}^{\rm i} }$\\
\hline
CLS$_{\log r_{\rm e}}$        &  0.22 &  -0.07  & -0.02 &  0.12 &  0.08 \\
CLS$_{< \! \mu \! >_{\rm e}}$ &  0.33 &   0.08  &  0.09 &  0.18 &  0.16 \\
CLS$_{\log n}$          & -0.35 &   0.08  & -0.01 & -0.16 & -0.03 \\
$\rm CORLS$                 & -0.10 &  -0.01  & -0.03 &  0.02 &  0.09 \\
$\rm CBLS$                  &  0.10 &   0.00  &  0.01 &  0.11 &  0.12 \\
$\rm CAMLS$                 &  0.02 &   0.03  &  0.03 &  0.09 &  0.12 \\
$\rm CGMLS$                 &  0.19 &   0.03  &  0.05 &  0.11 &  0.08 \\
\hline
\end{tabular}
\end{table}
We point out that  the correction procedure was repeated independently
for each  waveband of  \ms1008, without making  any assumption  {\it a
priori} on how  the properties of galaxies in  the space of structural
parameters can vary among the different wavebands.
\begin{table*}
\large
\caption{ Coefficients of the PHP in the $R$ band.}
\label{R_PHP}
\centering
\begin{tabular}{|l|c|c|c|c|c|}
\hline
& $a$ & $b$ & $c$ & $\sigma_{\log r_{\rm e}}$ &$\sigma_{\log r_{\rm e}}^{\rm i}$ \\
\hline
CLS$_{\log r_{\rm e}}$        &$    1.00 \pm    0.09$ &$   0.184 \pm   0.010$ &  $   -4.70 \pm    0.21$ & $    0.17 \pm    0.01 $ &$    0.15 \pm    0.01$ \\
CLS$_{< \! \mu \! >_{\rm e}}$ &$    0.90 \pm    0.14$ &$   0.303 \pm   0.019$ &  $   -7.07 \pm    0.38$ & $    0.21 \pm    0.02 $ &$    0.19 \pm    0.03$ \\
CLS$_{\log n}$          &$    1.40 \pm    0.16$ &$   0.157 \pm   0.025$ &  $   -4.22 \pm    0.40$ & $    0.18 \pm    0.02 $ &$    0.16 \pm    0.02$ \\
$\rm CORLS$                 &$    1.10 \pm    0.11$ &$   0.175 \pm   0.012$ &  $   -4.48 \pm    0.22$ & $    0.16 \pm    0.01 $ &$    0.14 \pm    0.01$ \\
$\rm CBLS$                  &$    1.07 \pm    0.06$ &$   0.219 \pm   0.009$ &  $   -5.41 \pm    0.19$ & $    0.17 \pm    0.01 $ &$    0.14 \pm    0.01$ \\
$\rm CAMLS$                 &$    1.09 \pm    0.08$ &$   0.215 \pm   0.010$ &  $   -5.34 \pm    0.20$ & $    0.17 \pm    0.01 $ &$    0.15 \pm    0.01$ \\
$\rm CGMLS$                 &$    1.07 \pm    0.09$ &$   0.201 \pm   0.017$ &  $   -5.04 \pm    0.34$ & $    0.17 \pm    0.01 $ &$    0.15 \pm    0.01$ \\
\hline
\end{tabular}
\end{table*}

\begin{figure*}
\centering
\includegraphics[angle=0,width=16cm,height=16cm]{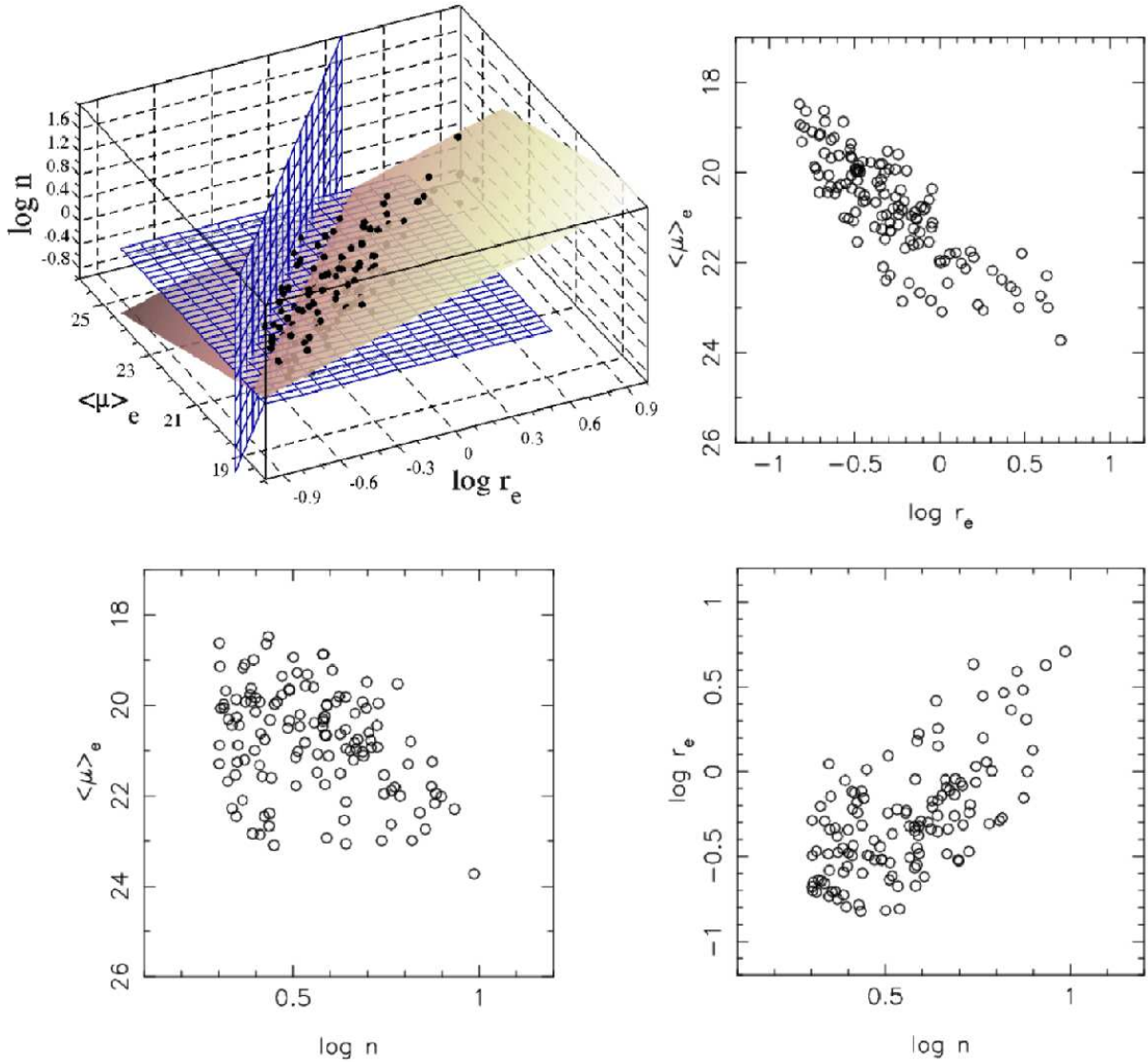}
  \caption{PHP  of \ms1008 \,  in the  R-band.  The  upper--left panel
  shows a 3D view of the  plane, where black circles mark positions of
  galaxies ($\rm  N=129$) in the space of  structural parameters.  The
  shaded surface is the PHP, as defined by the \bls \, fitting method,
  while the  vertical and horizontal  blue grids depict  the selection
  cuts  in magnitude and  Sersic index  respectively.  The  shading is
  realized  in  such  a  way  that  color  intensities  are  inversely
  proportional to \lsn.  Note that the plot of the PHP is transparent,
  and  therefore  black points  below  the  surface  appear as  partly
  obscured symbols.   The upper--right and  the lower panels  show the
  \lre--\mie,  \lsn--\mie  \, and  \lsn--\lre  \,  projections of  the
  PHP. }
  \label{PHP_3D}
\end{figure*}

\section{Photometric Planes of \ms1008}
\label{PHP_MS1008}
\begin{figure}
\centering
  \includegraphics[angle=0,width=8.0cm,height=8.0cm]{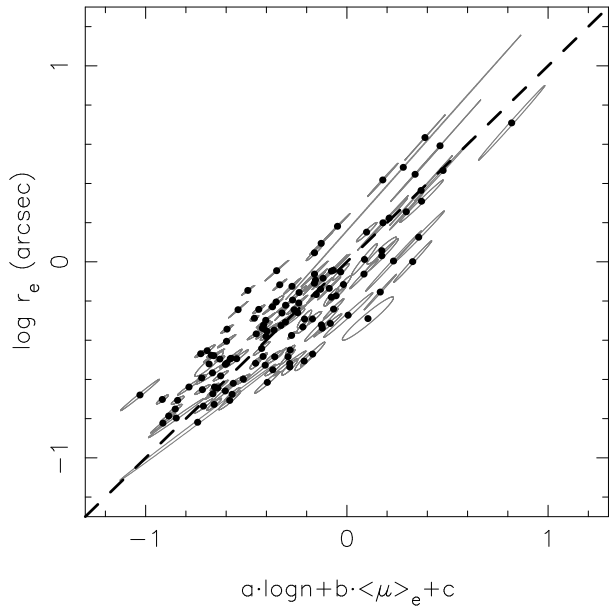}
  \caption{Edge-on  view of  the $R$-band  PHP of  \ms1008.   The plot
  shows the $N=129$ galaxies  of the $R$-band sample.  Ellipses denote
  uncertainties on structural parameters  and correspond to $1 \sigma$
  confidence contours.}
  \label{PHP_EDGE}
\end{figure}

The distribution of galaxies in  the space of structural parameters is
shown in Fig.~\ref{PHP_3D}, where we show the distribution of galaxies
in the  \lre--\mie, \lsn--\mie \, and  \lsn--\lre \, planes,  and a 3D
view of  the PHP,  with the corresponding  magnitude and  Sersic index
cuts represented by two orthogonal  planes.  An edge projection of the
plane  is also  shown in  Fig.~\ref{PHP_EDGE}.  We  see  that galaxies
follow  a  well-defined  PHP  at  $z=0.3$, with  Sersic  indices  that
increase towards lower surface  brightness values and larger effective
radii.

The coefficients  of the  PHP in the  $R$, $I$  and $K$ bands  for the
different      fitting       procedures      are      reported      in
Tables~\ref{R_PHP},~\ref{I_PHP} and~\ref{K_PHP} respectively, together
with the  estimate $\sigma_{\log r_{\rm e}}^{\rm i}$  of the intrinsic
dispersion  in  \lre  \,  about  the plane.   All  these  values  were
corrected for  selection effects  as detailed in  Section~\ref{SEL}.  The
$K$-band coefficients  were obtained by combining  the values obtained
for the \KS  and \KI \, bands, as  discussed below (Section~\ref{KKPHP}).
For the \cls  \, and \orls \, fits,  the quantity $\sigma_{\log r_{\rm
e}}^{\rm  i}$   was  estimated  by  projecting   $\sigma^{\rm  i}$  in
Eq.~\ref{CHIEQ} along the direction $\bmath \log r_{\rm e}$, while for
the other fitting procedures we subtracted in quadrature to the rms of
\lre \,  residuals the  amount of scatter  due to  measurement errors.
This  was  performed  by   taking  into  account  the  correlation  of
uncertaintities on \lre, \mie \, and \lsn.

By looking at  Tables~\ref{R_PHP},~\ref{I_PHP} and~\ref{K_PHP}, we see
that (i) the  coefficients of the PHP obtained  by the various fitting
procedures  are  significantly   different,  and  (ii),  whatever  the
regression method  is, the  PHP has significant  intrinsic dispersion.
As discussed by LBC00 for  the FP relation, the existence of intrinsic
scatter for a bivariate relation and our ignorance on its origin imply
that different  fitting methods do not  necessarily provide consistent
results.  We note, however, that the dependence of PHP coefficients on
the  fitting procedure is  particularly significant  only for  the CLS
regressions, which give a special role to the variable whose residuals
are minimized during the fit.  On the other hand, treating equally all
the variables, as in the CORLS, CBLS, CAMLS and CGMLS fits, gives much
more robust estimates, allowing us to obtain a stable determination of
PHP coefficients.  We  also note that the CLS$_{ <  \!  \mu \!  >_{\rm
e}}$ regression  provides a lower value  of $a$ and a  higher value of
$b$  with respect  to  the CLS$_{  \log  r_{\rm e}}$  method.  On  the
contrary, the CLS$_{\log n}$ regression produces a higher value of $a$
and a lower  value of $b$ with respect to the  CLS$_{ \log r_{\rm e}}$
fit.  Since the CBLS, CAMLS and  CGMLS fits are based on an average of
the CLS coefficients, the corresponding coefficients are very close to
those of the  CLS$_{ \log r_{\rm e}}$ method.   Since the coefficients
of the bisector fit have a smaller relative uncertainties with respect
to the other  fitting methods, we will refer  to these coefficients in
the following analysis  (Section~\ref{PHPZ0}).  It is also interesting
to note that the correction  for selection effects is negative for the
\orls \, fit while it is positive  for the \bls, \amls \, and \gmls \,
regressions, and that, therefore,  if selection effects would have not
taken into  account, the difference among  the \orls \,  and the \bls,
\amls \, and \gmls \, fits would have been significant.

\subsection{The PHP in optical wavebands}
\label{OPTPHP}
\begin{figure}
  \centering
  \includegraphics[angle=0,width=7.5cm,height=15cm]{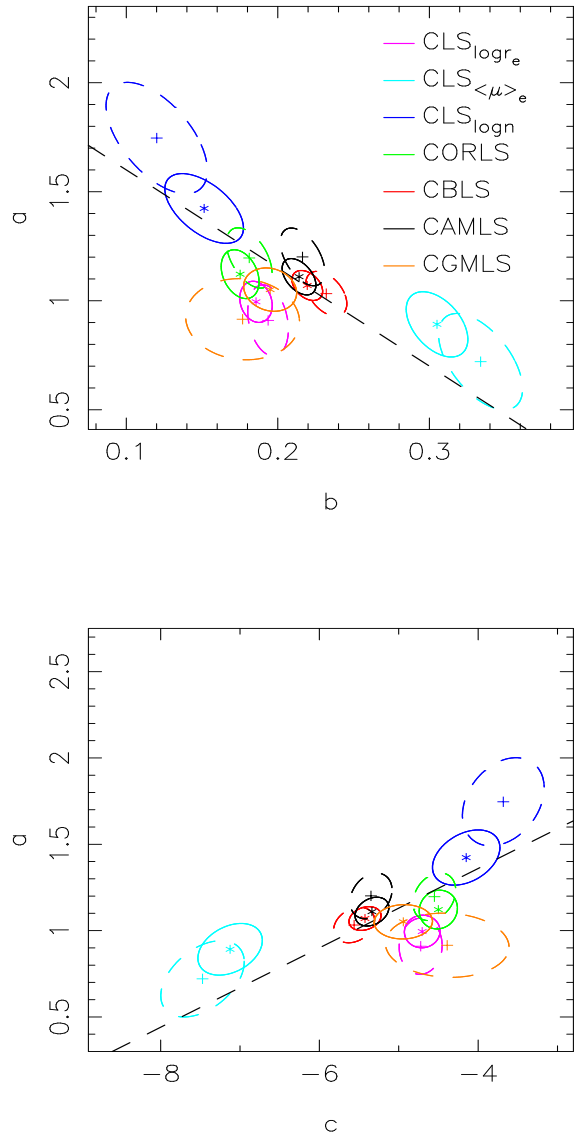}
  \caption{Comparison of  the optical PHP coefficients.   The $R$- and
  $I$-  band   coefficients  are  marked  by   asterisks  and  crosses
  respectively,  while the  fitting procedure  are shown  in different
  colors,  as  indicated  in  the  upper panel.   The  solid  (dashed)
  ellipses  mark  $1 \sigma$  confidence  contours  for  each pair  of
  R(I)-band coefficients  (see text).  The  dashed lines in  the upper
  and  lower panels  show the  direction  of the  mean correlation  of
  uncertainties on $a$ and $b$, and $b$ and $c$ respectively.}
  \label{PHPR}
\end{figure}

A proper comparison of  PHP coefficients between different samples has
to account for  the fact that their uncertainties  are correlated. For
this reason, we chose to compare simultaneously the slopes $a$ and $b$
of the PHP and one slope,  $a$, with its zeropoint $c$. The comparison
of $b$ versus $c$ as well as other plots with pairs of quantities from
Tables~\ref{R_PHP} and~\ref{I_PHP}  do not add  further information to
the   discussion   and  are   not   shown   in   the  following.    In
Fig.~\ref{PHPR}, we plot the  $R$- and $I$- band coefficients.  Taking
into account  the mean galaxy colour,  $\overline{R \!  -  \!  I} \sim
0.75$ at $z  \sim 0.3$, we expect an offset between  the $I$- and $R$-
band PHP zero-points  of $\Delta_{R-I} \sim b \cdot  \overline{R \!  -
\!  I}  $.  In  order to  remove this effect  from the  comparison, we
subtracted the quantity $\Delta_{R-I}$ from  the I- band values of $c$
which are shown  in Fig.~\ref{PHPR}.  We note that  since the value of
$\Delta_{R-I}$  is  proportional to  $b  \sim  0.2$, uncertainties  on
$\overline{R  \!   -  \!   I}  $  do  not  affect  the  comparison  in
Fig.~\ref{PHPR}.

As  shown in  the figure,  the PHP  coefficients are  fully consistent
between the $R$  and $I$ bands. This is clearly  in agreement with the
fact  that structural  parameters do  not show  significant variations
among optical wavebands (see  Appendix~\ref{UNCER}), and is due to the
fact  that the  $R$ and  $I$ bands  at $z=0.3$  sample a  very similar
spectral region, where differences in stellar population properties of
galaxies,  such as  their  internal colour  gradients  and the  colour
magnitude relation, are negligible.   For what concerns the dispersion
around the plane, we  see from Tables~\ref{R_PHP} and~\ref{I_PHP} that
this is  also consistent between the  $R$ and $I$  bands, amounting on
average to  $\sim 0.18$~dex  ($\sim 41 \%$)  in \lre. The  same result
holds for the intrinsic dispersion  of the PHP, which amounts to $\sim
0.14-0.15$~dex ($\sim 32-35  \%$) in \lre.  We note  that although for
the CLS$_{< \!  \mu \!  >_{\rm  e}}$ and CLS$_{\log n}$ fits the value
of  \sir \,  is slightly  larger  with respect  to that  of the  other
fitting  methods,  the  corresponding  uncertainties of  \sir  \,  are
larger, making this difference  not particularly significant.  We also
note that only a few percent  of the observed dispersion about the PHP
are explained by the the measurement errors on \lre, \mie \, and \lsn.
As shown in Fig.~\ref{PHP_EDGE},  in fact, uncertainties on structural
parameters  are strongly  correlated in  a direction  which  is almost
parallel to the PHP.

We  also  analyzed  the  presence  of  a  possible  curvature  in  the
distribution  of ETGs  in the  space  of \lre,  \mie \,  and \lsn,  by
considering correlations  among \lre \, residuals about  the plane and
each of the three variables.  We found that such correlations are very
weak,  the Spearman's  rank  coefficients amounting  to  0.14 for  the
correlation  between residuals  and \lre,  and to  -0.2 for  the other
correlations with \mie  \, and \lsn. These coefficients  can be easily
explained by the selection cuts in the space of structural parameters.
We also fitted the $R$-band PHP by using \lre \, as dependent variable
and by adding to Eq.~\ref{PHPEQ}  one among all the possible quadratic
terms which  can be  constructed from the  two quantities \mie  \, and
\lsn.  All  these terms were found  to be consistent with  zero at the
level of  $1.4 \sigma$, the \lre  \, scatter of the  PHP decreasing by
less than  $1 \%$  in each case.   {\it Our data,  therefore, indicate
that  there  is  no significant  departure  of  the  PHP from  a  flat
relation.}

\begin{figure}
  \centering
  \includegraphics[angle=0,width=7.5cm,height=15cm]{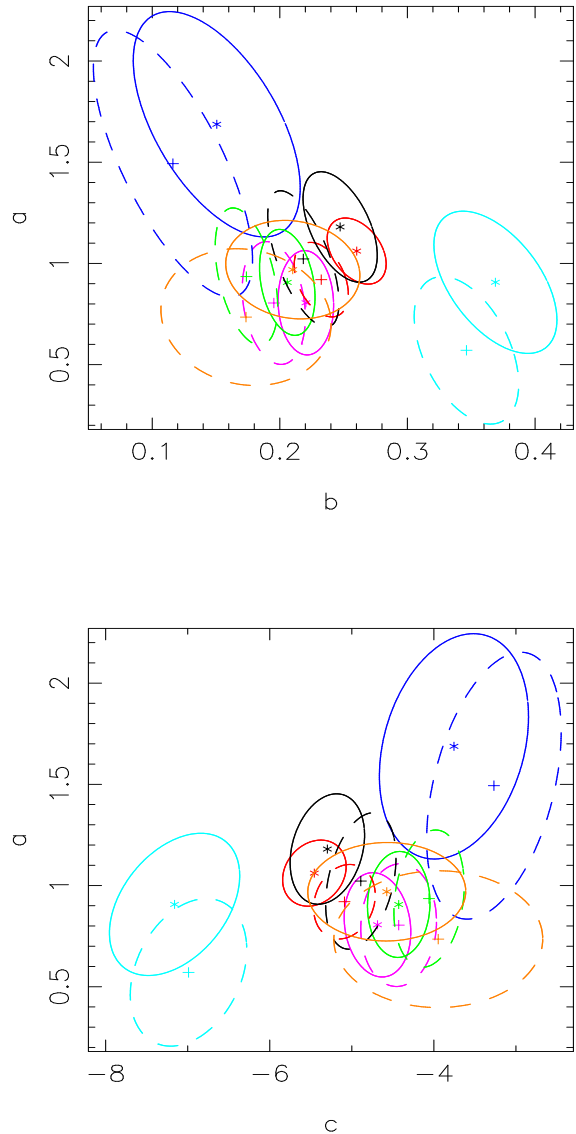}
  \caption{The  same of  Fig.~\ref{PHPR} for  the  \KS \,  and \KI  \,
  coefficients.  Asterisks and crosses mark  the \KS \, and \KI \, PHP
  coefficients respectively.  Ellipses  denote $1.5 \sigma$ confidence
  contours.  }
  \label{PHPKK}
\end{figure}
\begin{table*}
\large
\caption{ Coefficients of the PHP in the $I$ band.}
\label{I_PHP}
\centering
\begin{tabular}{|l|c|c|c|c|c|}
\hline
& $a$ & $b$ & $c$ & $\sigma_{\log r_{\rm e}}$ & $\sigma_{\log r_{\rm e}}^{\rm i}$ \\
\hline
CLS$_{\log r_{\rm e}}$        &$    0.90 \pm    0.15$ &$   0.194 \pm   0.013$ &  $   -4.73 \pm    0.25$ & $    0.18 \pm    0.01 $ &$    0.14 \pm    0.01$ \\
CLS$_{< \! \mu \! >_{\rm e}}$ &$    0.75 \pm    0.20$ &$   0.331 \pm   0.026$ &  $   -7.32 \pm    0.51$ & $    0.22 \pm    0.02 $ &$    0.20 \pm    0.02$ \\
CLS$_{\log n}$          &$    1.66 \pm    0.24$ &$   0.124 \pm   0.029$ &  $   -3.70 \pm    0.46$ & $    0.25 \pm    0.04 $ &$    0.10 \pm    0.05$ \\
$\rm CORLS$                 &$    1.19 \pm    0.13$ &$   0.182 \pm   0.015$ &  $   -4.53 \pm    0.26$ & $    0.19 \pm    0.02 $ &$    0.14 \pm    0.01$ \\
$\rm CBLS$                  &$    1.03 \pm    0.09$ &$   0.232 \pm   0.013$ &  $   -5.52 \pm    0.24$ & $    0.19 \pm    0.01 $ &$    0.15 \pm    0.01$ \\
$\rm CAMLS$                 &$    1.17 \pm    0.12$ &$   0.218 \pm   0.014$ &  $   -5.33 \pm    0.24$ & $    0.20 \pm    0.01 $ &$    0.16 \pm    0.01$ \\
$\rm CGMLS$                 &$    0.95 \pm    0.16$ &$   0.186 \pm   0.025$ &  $   -4.55 \pm    0.51$ & $    0.19 \pm    0.02 $ &$    0.15 \pm    0.02$ \\
\hline
\end{tabular}
\end{table*}

\begin{table*}
\large
\caption{ Coefficients of the PHP in the $K$ band.}
\label{K_PHP}
\centering
\begin{tabular}{|l|c|c|c|c|c|}
\hline
& $a$ & $b$ & $c$ & $\sigma_{\log r_{\rm e}}$ & $\sigma_{\log r_{\rm e}}^{\rm i}$ \\
\hline
CLS$_{\log r_{\rm e}}$        & $    0.83 \pm    0.13$ &$   0.209 \pm   0.010$ & $   -4.51 \pm    0.19$ & $    0.18 \pm    0.01 $ &$    0.15 \pm    0.02$ \\
CLS$_{< \! \mu \! >_{\rm e}}$ & $    0.75 \pm    0.17$ &$   0.351 \pm   0.020$ & $   -6.83 \pm    0.32$ & $    0.21 \pm    0.02 $ &$    0.17 \pm    0.02$ \\
CLS$_{\log n}$          & $    1.39 \pm    0.29$ &$   0.151 \pm   0.033$ & $   -3.51 \pm    0.39$ & $    0.27 \pm    0.05 $ &$    0.23 \pm    0.05$ \\
$\rm CORLS$                 & $    0.90 \pm    0.14$ &$   0.192 \pm   0.010$ & $   -4.18 \pm    0.19$ & $    0.19 \pm    0.01 $ &$    0.17 \pm    0.02$ \\
$\rm CBLS$                  & $    0.97 \pm    0.08$ &$   0.245 \pm   0.010$ & $   -5.13 \pm    0.17$ & $    0.20 \pm    0.01 $ &$    0.15 \pm    0.02$ \\
$\rm CAMLS$                 & $    1.00 \pm    0.15$ &$   0.235 \pm   0.013$ & $   -4.98 \pm    0.19$ & $    0.21 \pm    0.02 $ &$    0.19 \pm    0.02$ \\
$\rm CGMLS$                 & $    0.91 \pm    0.13$ &$   0.212 \pm   0.027$ & $   -4.57 \pm    0.49$ & $    0.19 \pm    0.02 $ &$    0.16 \pm    0.02$ \\
\hline
\end{tabular}
\end{table*}

\subsection{The $K$-band plane}
\label{KKPHP}
In order to  derive the PHP of  \ms1008 \, in the $K$  band we applied
separately the different  fitting procedures to the \KS  \, and \KI \,
structural  parameters.  The  comparison  of  the \KS  \,  and \KI  \,
coefficients is  shown in Fig.~\ref{PHPKK}, where  the same quantities
as Fig.~\ref{PHPR} are  shown.  The important outcome is  that the PHP
coefficients  of  the  NIR   samples  are  fully  consistent,  with  a
confidence level  $\sol 1.5  \sigma$.  The same  result holds  for the
dispersion around  the plane.  This  is a reassuring result  since the
\KS \, and \KI \,  parameters were obtained from images with different
resolution and seeing.

Since the coefficients  and dispersions of the \KS \,  and \KI \, PHPs
turned out  to be consistent, we  combined their values  by a weighted
mean. The resulting coefficients are shown in Table~\ref{K_PHP}.

\section{PHP dependence  on waveband and redshift}

\subsection{The PHP in optical and NIR wavebands}
\label{OPTKKPHP}
\begin{figure}
  \centering
  \includegraphics[angle=0,width=8cm,height=16cm]{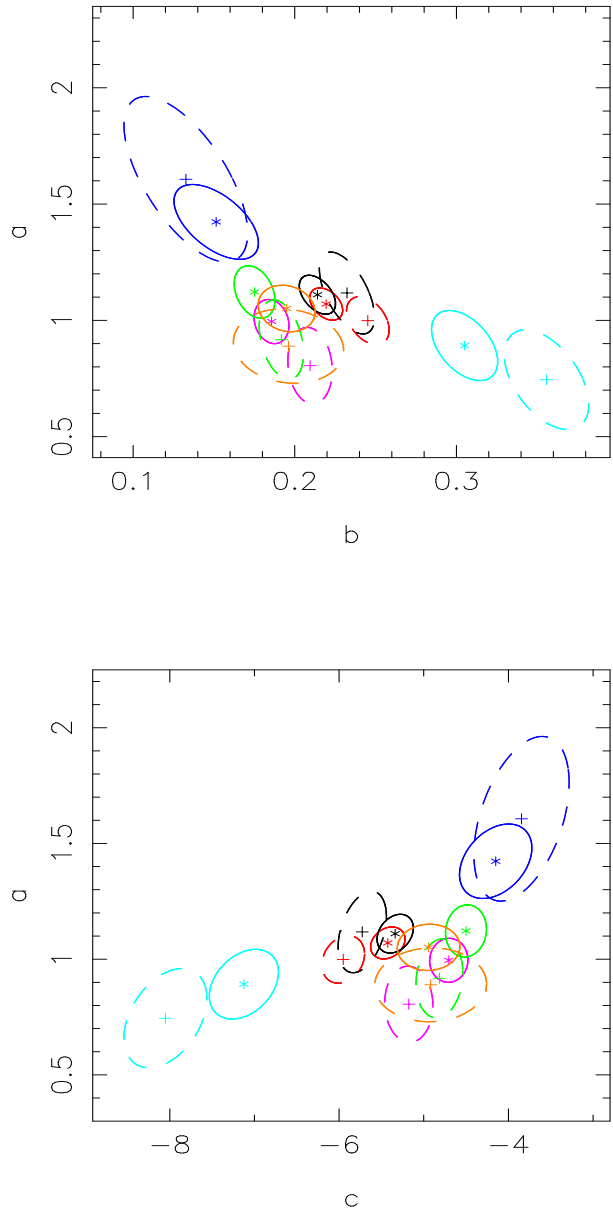}
  \caption{Comparison  of the  PHP coefficients  for the  $R$  and $K$
  bands (asterisks and crosses respectively).  Solid (dashed) ellipses
  denote $1 \sigma$ confidence contours for R(K)-band coefficients.
}
  \label{PHPRK}
\end{figure}

The  $R$-  and  $K$-  band  coefficients  of  the  PHP  are  shown  in
Fig.~\ref{PHPRK}.  In order to remove  the mean galaxy colour from the
comparison, we  proceeded as in Section~\ref{OPTPHP}  by subtracting from
the coefficient $c$ in the $K$  band the term $\Delta_{R-K} = -b \cdot
\overline{R \!  -  \!  K}$, with $\overline{R-K} \sim  3.2$ at $z \sim
0.3$.
Since the $I$-band  coefficients are consistent with those  in the $R$
band, they  do not add further  information to the  discussion and are
not  shown in  the  figure.   By looking  at  Fig.~\ref{PHPRK} and  at
Tables~\ref{R_PHP},~\ref{I_PHP} and~\ref{K_PHP}, we see that, whatever
regression method is adopted, {\it the PHP coefficients turn out to be
fully consistent among  the optical and NIR wavebands}.   We note that
since  the Sersic  index and  the magnitude  cuts affect  each fitting
procedure in a different way,  the independence of the above result on
the  fitting method  makes it  very robust  with respect  to selection
effects  in the  samples. Considering  the \orls,  \bls, \amls  \, and
\gmls  \, fits, we  find $a  \sim 1  $ and  $b \sim  0.2$ both  in the
optical and  NIR wavebands.  We  also note that the  intrinsic scatter
about the $K$-band  plane is fully consistent with  the value of $\sim
0.14-0.15$~dex found  for the  optical PHP, provided  that measurement
uncertainties are taken into account.

\subsection{The PHP at $z \sim 0.3$ and $z \sim 0$}
\label{PHPZ0}
\begin{figure}
  \centering
  \includegraphics[angle=0,width=8cm,height=8cm]{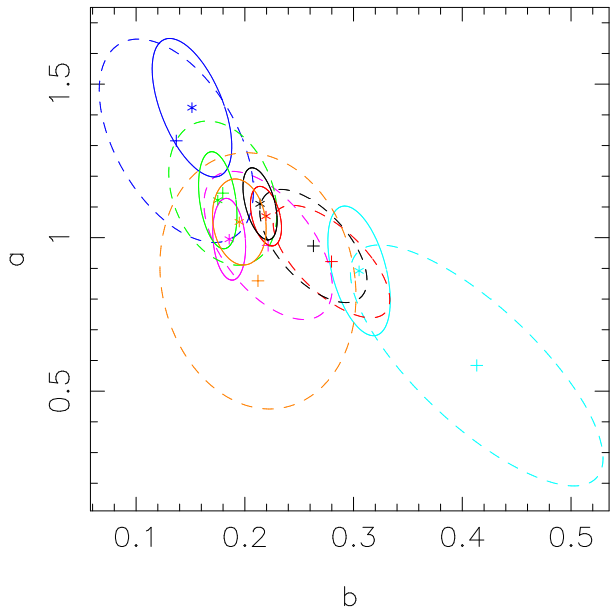}
  \caption{Comparison  of  the  $R$-band  PHP  slopes  of  \ms1008  \,
  (crosses) with those  of nearby galaxies in the  $B$ band from GRA02
  (asterisks).  Ellipses denote $1.5 \sigma$ confidence contours.  }
  \label{CONF_R_GRA02}
\end{figure}
In order  to address the redshift  dependence of the  PHP, we compared
our results with those of GRA02, who  derived the PHP at $z \sim 0$ by
using $B$-band structural parameters for  $N=38$ ETGs in the Virgo and
Fornax clusters.   We note that the  $R$ band at $z=0.3$  is closed to
$B$-band  restframe, and  therefore the  data  of GRA02  and those  of
\ms1008  \, cover  approximately the  same restframe  wavelengths.  To
perform a homogeneous comparison, we used the data in table~1 of GRA02
and  re-derived  the PHP  coefficients  using  the regression  methods
described  in Section~\ref{FITS}.  Selection  effects were  taken into
account   by   constructing    PHP   simulations   as   described   in
Section~\ref{SEL} with the same magnitude and Sersic index cuts as the
GRA02 sample,  i.e.  $B=15$ and $n=1.5$, respectively.   We note that,
since the  GRA02 sample is  not complete in  any sense (see  sec.~2 of
GRA02), the  above procedure provides  only a rough correction  of the
selection biases.   Because of the  wide magnitude range of  the GRA02
sample ($\sim 5.3$~mag), the corrections turned out to be quite small,
amounting  to $5  \%$  on average,  and to  at  most $10  \%$ for  the
coefficient $a$  in the  CLS$_{\log n}$ fit.   The bias  corrected PHP
coefficients at  $z \sim 0$  are given in  Table~\ref{GRA02_PHP}.  The
intrinsic  dispersion  of the  PHP  was  computed  assuming a  typical
uncertainty of  $\sim 0.1$~dex in \lsn  \, (see sec.~2  of GRA02), and
adopting  as mean  covariance  matrix of  uncertainties on  structural
parameters that of  galaxies in \ms1008, which was  re-scaled to match
the  uncertainty on \lsn  \, at  $z \sim  0$.  The  slopes of  the PHP
obtained  by the  \orls,  \bls, \amls  \,  and \gmls  \,  fits can  be
compared with the values of $a=0.89 \pm 0.14$ and $ b=0.24 \pm 0.036$,
that were derived from GRA02 by the bisector fit, treating equally all
the three  variables.  The  values in Table~\ref{GRA02_PHP}  are fully
consistent with  those of GRA02.  The  PHP slopes of  the GRA02 sample
and those of \ms1008 \, are compared in Fig.~\ref{CONF_R_GRA02}, where
we see that the  values of a and b at $z  \sim 0$ are fully consistent
with those  at $z \sim 0.3$.  We  note that due to  the smaller sample
size at  $ z \sim  0$ the corresponding  uncertainties on a and  b are
larger,  particularly for some  of the  fitting procedures  (e.g.  the
CLS$_{<   \!     \mu   \!    >_{\rm   e}}$   fit).     As   shown   in
Table~\ref{GRA02_PHP}, the intrinsic  dispersion around the plane does
not change significantly,  provided that measurement uncertainties are
taken into account.
\begin{table*}
\large
\caption{ $B$-band coefficients of the PHP for the GRA02 sample.}
\label{GRA02_PHP}
\centering
\begin{tabular}{|l|c|c|c|c|c|}
\hline
& $a$ & $b$ & $c$ & $\sigma_{\log r_{\rm e}}$ & $\sigma_{\log r_{\rm e}}^{\rm i}$ \\
\hline
CLS$_{\log r_{\rm e}}$        &$    0.97 \pm    0.16$ &$   0.221 \pm   0.039$ &  $   -4.87 \pm    0.80$ & $    0.17 \pm    0.01$ & $    0.14 \pm    0.02 $\\
CLS$_{< \! \mu \! >_{\rm e}}$ &$    0.58 \pm    0.27$ &$   0.433 \pm   0.078$ &  $   -9.00 \pm    1.55$ & $    0.20 \pm    0.04$ & $    0.17 \pm    0.04 $\\
CLS$_{\log n}$          &$    1.31 \pm    0.24$ &$   0.135 \pm   0.052$ &  $   -3.18 \pm    0.98$ & $    0.21 \pm    0.04$ & $    0.15 \pm    0.03 $\\
$\rm CORLS$                 &$    1.14 \pm    0.15$ &$   0.180 \pm   0.031$ &  $   -4.07 \pm    0.64$ & $    0.19 \pm    0.02$ & $    0.14 \pm    0.01 $\\
$\rm CBLS$                  &$    0.92 \pm    0.12$ &$   0.279 \pm   0.036$ &  $   -5.92 \pm    0.71$ & $    0.17 \pm    0.01$ & $    0.13 \pm    0.01 $\\
$\rm CAMLS$                 &$    0.97 \pm    0.12$ &$   0.263 \pm   0.034$ &  $   -5.58 \pm    0.66$ & $    0.17 \pm    0.01$ & $    0.13 \pm    0.01 $\\
$\rm CGMLS$                 &$    0.85 \pm    0.30$ &$   0.212 \pm   0.060$ &  $   -4.45 \pm    1.25$ & $    0.18 \pm    0.05$ & $    0.16 \pm    0.06 $\\
\hline

\end{tabular}
\end{table*}

\begin{figure}
  \centering
  \includegraphics[angle=0,width=8cm,height=8cm]{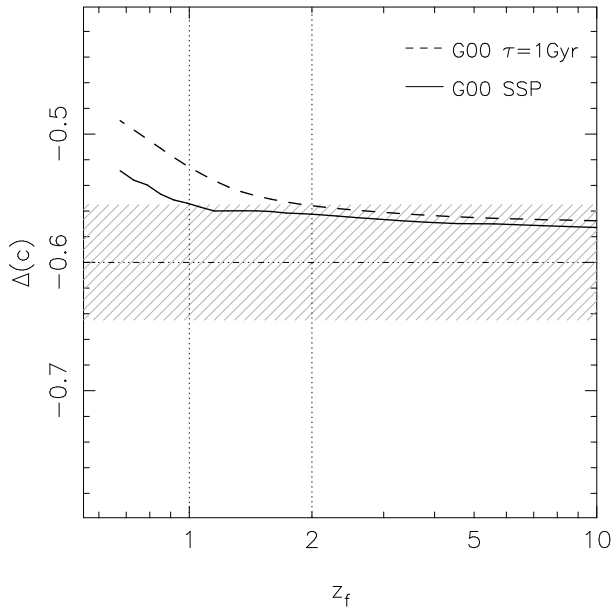}
  \caption{The difference of the PHP zero-point, $\Delta (c)$, between
  $z=0.3$ and $z=0$ is compared with predictions of stellar population
  models.  The  dash-dotted line and the gray  hatched region indicate
  the  mean value  of  $\Delta  (c)$ and  the  corresponding $\pm  1.5
  \sigma$   confidence   interval.    The   dotted  lines   mark   the
  intersections of the two models with the hatched area.}
  \label{PHP_ZP_EV}
\end{figure}

Taking  advantage of  the  fact  that the  PHP  slopes are  consistent
between $z  \sim 0$ and $z  \sim 0.3$, we  re-computed the coefficient
$c$ at $z \sim 0$ and $z \sim  0.3$ by fixing the slopes of the PHP to
the values of $a$ and $ b$ obtained for \ms1008 \, by the \bls \, fit,
i.e.  $a=1.07$  and $b=0.219$. The \bls  \, fit was  chosen because of
the  smaller uncertainties  of bisector  coefficients with  respect to
those of other fitting procedures.  The value of $c$ at $z \sim 0$ was
also corrected  for magnitude  and Sersic index  cuts as  described in
Section~\ref{SEL}.  Due to the robustness  of the \bls \, regression with
respect to selection effects (see Table~\ref{CORPHPR}), the correction
turned  out to be  very small  ($\sim 0.2  \%$).  The  coefficient $c$
amounts to $-4.737 \pm 0.028$ at $z \sim 0$ and $-5.337 \pm 0.014$ for
\ms1008, differing  by $\Delta  (c) = -0.60  \pm 0.03$.  The  value of
$\Delta (c)$ can  be analyzed by taking the  average of the difference
between the  PHP equations (Eq.~\ref{PHPEQ}) at  $z \sim 0.3$  and $ z
\sim 0$:
\begin{equation}
\Delta (c)  =  \Delta  (\overline{\log  r_{\rm e}})  -  a  \cdot  \Delta
(\overline{\log n}) - b \cdot \Delta ( \overline{< \! \mu \! >_{\rm e}} ).
\label{DELTAC_1}
\end{equation}
Assuming that, on average, structural  parameters vary from $z \sim 0$
 to $z \sim  0.3$ only because of the  luminosity evolution of stellar
 populations, we can re-write the previous equation by setting $\Delta
 (\overline{\log  r_{\rm e}})=0$  and  $\Delta (\overline{\log  n})=0$
 (see discussion in Section~\ref{SD3}):
\begin{equation}
\Delta  (c) = - \log(d_{\rm A}) -  b \cdot \left[ 10  \Delta \left[ \log
    \left(  \frac{1+z}{1.008} \right)  \right] -  (B_{0.0}  - R_{0.3})
    \right],
\label{DELTAC}
\end{equation}
where the term $10 \Delta \left[ \log \left( \frac{1+z}{1.008} \right)
\right]$  accounts  for the  surface  brightness  dimming between  the
redshift  of \ms1008  and that  of the  GRA02  sample\footnote{For the
GRA02  sample of Virgo  and Fornax  galaxies, we  assumed $z=0.008$.},
$d_{\rm  A}$ is  the  angular diameter  distance\footnote{The term  $-
\log(d_{\rm  A})$  accounts  for  the  fact that  effective  radii  of
galaxies in \ms1008 \, are given in arcsec, while for the GRA02 sample
\re \, is given in kpc.}  corresponding to $z =0.306$, while $(B_{0.0}
- R_{0.3})$ is the difference between  the $B$ magnitude at $z \sim 0$
and the $R$-band magnitude at $z \sim 0.3$.  Eq.~\ref{DELTAC} was used
to  compare the  measured value  of $\Delta  (c)$ with  predictions of
stellar  population models from  the GISSEL03  code~\citep{BrC03}.  We
considered  two   different  stellar  population   models  with  solar
metallicity  and a  Scalo IMF  \citep{SCALO},  the first  one being  a
simple   stellar  population   (SSP),   the  second   one  having   an
exponentially declining  star formation rate  with time scale  $\tau =
1~Gyr$.  For  both models, we  computed the value of  $\Delta(c)$ from
Eq.~\ref{DELTAC}  by deriving  the $B$-  and $R$-  band  magnitudes at
$z=0$  and $z=0.306$ respectively,  for different  formation redshifts
$z_{\rm  f}$.    In  Fig.~\ref{PHP_ZP_EV},  we  show   the  values  of
$\Delta(c)$ for the  two stellar populations as a  function of $z_{\rm
f}$.  The measured value of  $\Delta(c)$ is marked by the dash--dotted
line in  the plot, while the  hatched area is  the confidence interval
defined by measurement uncertainties on $\Delta(c)$.  The figure shows
that  both  models  cross  the  hatched  region  and  are  consistent,
therefore,  with  the measured  value  of  $\Delta(c)$.  The  crossing
points  define  lower limits  for  the  formation  redshift of  galaxy
stellar populations.  For the SSP and $\tau = 1~Gyr$ models, we obtain
$z_{\rm f} > 1$ and $ z_{\rm f} > 2$ respectively.

%
%
%
%
%
%

\section{Discussion}
\label{DISC}
We have  shown that  cluster ETGs at  $z \sim  0.3$ \, follow  a tight
correlation among \lre, \mie \, and \lsn, with an intrinsic dispersion
of  $\sim 32  \%$  in \re.   Our data  indicate  that the  PHP has  no
significant curvature.  In Section~\ref{D0} we discuss whether stellar
populations can  be the origin  of the intrinsic dispersion  about the
plane,  while  in  Section~\ref{D1}  we  attempt to  find  a  possible
explanation    for     the    absence    of     curvature    in    the
PHP.  Sections~\ref{SD2}  and~\ref{SD3}  deal  with the  waveband  and
redshift dependence of the PHP coefficients.

\subsection{The intrinsic dispersion of the PHP}
\label{D0}
\begin{figure}
  \centering
  \includegraphics[angle=0,width=8cm,height=8cm]{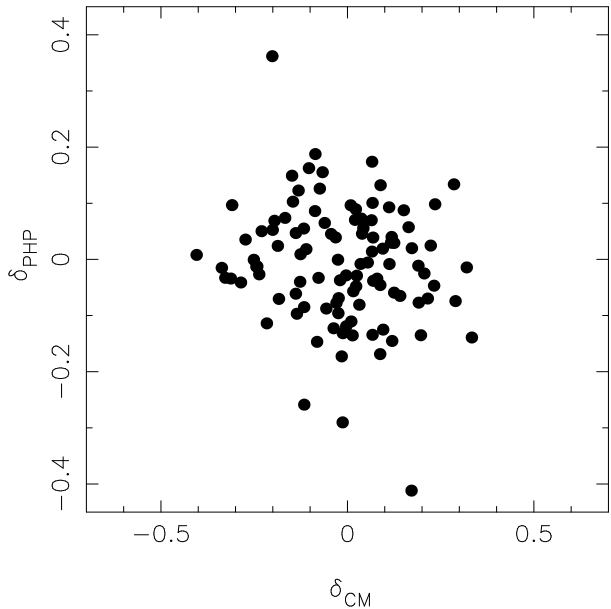}
  \caption{ Relation among residuals to the CM and PHP relations.  }
  \label{resid}
\end{figure}

In  Section~\ref{PHP_MS1008}  we  showed  that the  PHP  relation  has
significant intrinsic  dispersion, amounting to  $\sigma^{\rm i}_{\log
r_{\rm  e}} \sim  0.14$~dex ($\sim  32 \%$)  both in  optical  and NIR
wavebands.   This dispersion is  fully consistent  with that  found by
GRA02 for ETGs  in nearby clusters.  The intrinsic  scatter of the PHP
turns out, therefore, to be larger with respect to that of the FP (see
values reported  in Section~\ref{intr}),  although it is  smaller with
respect to  that of other  correlations among galaxy  parameters.  For
example,   \citet{PDdC98}   analyzed   different  correlations   among
photometric and  spectroscopic properties of nearby  ETGs.  They found
that the Kormendy  relation (KR) and the $\rm Mg_2$ FP  have a \lre \,
scatter of $\sim 0.23$~dex and $\sim 0.17$~dex respectively, where the
$\rm  Mg_2$ FP  relation  was constructed  by  replacing the  velocity
dispersion term in  the FP with the $\rm  Mg_2$ line--strength.  It is
interesting  to compare  these values  with  those we  can obtain  for
\ms1008.  A KR  fit to the $R$-band sample of \ms1008  \, gives a \lre
\, dispersion of $\sim 0.2$~dex, that  is $\sim 14 \%$ higher than the
scatter  of the PHP.   This result  is consistent  with what  found by
LMB04. Since we do not have $\rm Mg_2$ line--strengths, we constructed
a `photometric' FP by  replacing velocity dispersions with optical-NIR
galaxy  colours (see  e.g.~\citealt{dCD89}).  Using  $I \!   -  \!  K$
colours  available for $N=105$  galaxies from  the $R$-band  sample of
\ms1008, we obtain $\log r_{\rm e}  \propto (0.25 \pm 0.02) \cdot < \!
\mu \!   >_{\rm e} + (0.23  \pm 0.07) \cdot (I  \!  - \!   K)$, with a
\lre \, rms  of $0.175$~dex, which is $\sim 8  \%$ larger with respect
to the PHP,  but significantly smaller with respect  to the KR.  Since
effective  parameters turn  out  to correlate  with  both colours  and
Sersic indices,  we also tried  to construct a  photometric hyperplane
with \lre, \mie,  \lsn \, and \,  $I \!  - \! K$.   We found, however,
that the $I \!  - \!  K$  term in such a hyperplane is only marginally
significant ($1.5  \sigma$), and that  the scatter about  the relation
decreases only by $1\%$ with respect  to that of the PHP.  This result
is also  shown in  Fig.~\ref{resid}, where we  plot \lre  \, residuals
about the PHP  versus residuals from the  $I \!  - \!  K$  vs.  $K$ CM
relation of \ms1008.  Such  $\delta-\delta$ diagrams have been already
applied in several works to  investigate the origin of residuals about
the FP  (e.g.~\citealt{PrS96}).  The plot clearly shows  that there is
no  correlation among  residuals  to  the PHP  and  CM relations,  the
corresponding  Spearman's  rank coefficient  amounting  to $-0.12  \pm
0.09$.   This implies  that either  the scatter  of the  PHP  does not
origin from  stellar population parameters or that  a complex combined
effect of  such parameters  (e.g.  age and  metallicity) is  acting in
such  a way  that  no correlation  appears  in the  $\delta$--$\delta$
diagram.  This  issue could be addressed by  correlating PHP residuals
with line--strength indices.

\subsection{Why a plane with \lre, \mie \, and \lsn ?}
\label{D1}
As mentioned  in Section~\ref{intr}, MLC00  and MLC01 found  that ETGs
populate  a curved  manifold in  the space  of  structural parameters,
showing  that the  existence of  such a  PHP can  be explained  by the
specific entropy of  ETGs, $s$, being an increasing  function of their
mass, $M$.   MLC01 also showed  that the existence  of a plane  in the
space of  \lre, \mi0 \,  and \lsn \,  can be explained by  the limited
range considered  for $\log n$ in  previous works (with $  \log n \sog
0.15$, see  fig.~10 of MLC01).  Since  the variables used  by MLC01 as
well as  the quantities  \lre, \mi0  \, and \lsn  \, are  not linearly
related  to \lre,  \mie \,  and \lsn  \, (see  Section~\ref{intr}), an
intriguing question  arises on what the  origin is of  the flatness of
the PHP in the \lre, \mie \, and \lsn \, space.

\begin{figure}
  \centering
  \includegraphics[angle=0,width=8cm,height=8cm]{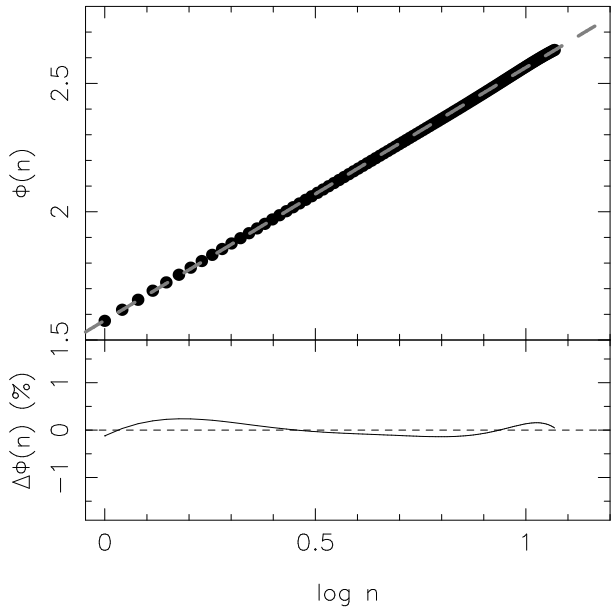}
  \caption{Relation between  the function $\Phi(n)$  and the logarithm
  of the  Sersic index (see  the text).  The  grey curve in  the upper
  panel marks the best-fitting line  of $\Phi(n)$ vs. \lsn.  The lower
  panel shows the residuals of the linear fit (solid line). }
  \label{ENTROP1}
\end{figure}

To address this issue, we tried  to understand if there is some simple
relation between the specific entropy of ETGs and the three quantities
\lre,  \mie \, and  \lsn. This  was achieved  by considering  the same
galaxy  models  described  in  MLC00  and MLC01,  that  is  spherical,
non-rotating  systems,   with  negligible  radial   gradients  of  the
mass-to-light ratio  (i.e.  one component  models). We point  out that
attempting to derive an interpretation for the existence of the PHP on
the  basis of  more complex  models is  well beyond  the scope  of the
present work, and  therefore we do not discuss  further here the above
mentioned hypotheses.   As detailed in Appendix~\ref{AENT},  it can be
shown that the specific entropy is given by the following formula:
\begin{equation}
\begin{array}{lll}
s & =  & 0.5 \ln \! 10 \cdot \log  M/L + 2.5 \ln \! 10 \cdot \log  r_{\rm e} + \\ 
& - & 0.2 \ln \! 10 < \! \mu \! >_{\rm e} + \Phi(n) + const.
\label{SEEQ2}
\end{array}
\end{equation}
where $ \Phi(n)$ is  an dimensionless function given by:
\begin{equation}
\begin{array}{lll}
\Phi(n) & =  & 0.5 \ln \left( \frac{b_n^{2n}}{n  \Gamma(2n)} \right) +
        \\  &   +  &   2  \frac{b_n^{2n}}{n  \Gamma(2n)}   \cdot  \int
        \tilde{\rho}  \ln  \left( \tilde{P}^{3/2}  \tilde{\rho}^{-5/2}
        \right) d \! s,
\label{SEEQ3}
\end{array}
\end{equation}
$ \tilde{P}$  and $ \tilde{\rho}$ being the  dimensionless 3D pressure
and density profiles of the  models respectively. Since $\Phi(n)$ is a
very complex function of the Sersic  index, there is no {\it a priori}
reason  for which  a  systematic  variation of  $s$  along the  galaxy
sequence (such as  the $s$--$M$ relation found by  MLC00) should imply
the existence of a plane in the space of \lre, \mie \, and \lsn.  From
Eq.~\ref{SEEQ2}, we  see that this  is the case  only and only  if the
function   $\Phi(n)$    is   a   linear   function    of   \lsn.    In
Fig.~\ref{ENTROP1} we plot the function $\Phi(n)$ for different values
of \lsn.  The values of  $\Phi(n)$ were computed numerically, by using
the formulae in Appendix~\ref{AENT}.  By  looking at the figure, it is
remarkable to note  that there exists almost an  exact linear relation
between  $\Phi(n)$ and \lsn.   A linear  best-fit gives  $\Phi(n) \sim
0.988 \cdot \log (n) + 1.57$,  with residuals smaller than $0.5 \%$ in
absolute value. We  point out that the existence  of a linear relation
between $\Phi(n)$  and \lsn \,  holds in a  wide range of  $n$ values,
from $n  \sim 1$ to $n  \sim 10$, and it  is not a  consequence of the
limited  range   considered  for  the  Sersic   index.   We  conclude,
therefore, that the flatness of  the PHP of ETGs is strictly connected
to the physical origin of  this relation, the specific entropy of ETGs
being a  linear combination of the  variables \lre, \mie  \, and \lsn.
In the future,  it will be very interesting  to derive Eq.~\ref{SEEQ2}
by considering more complex galaxy models (e.g. by including both dark
and luminous  matter) and to  analyze the implications of  such models
for  the  existence of  a  flat surface  in  the  space of  structural
parameters.

\subsection{The waveband dependence of the PHP}
\label{SD2}
Since  the  variation  of  \mie  \,  between  different  wavebands  is
proportional to the galaxy colour (see Eq.\ref{magn}) and therefore to
galaxy luminosity, through the CM  relation, the PHP slopes of \ms1008
\, in the  optical and NIR wavebands can be used  to constrain how the
ratios  $\frac{n_{\rm R}}{n_{\rm  K}}$ and  $\frac{r_{\rm e,R}}{r_{\rm
e,K}}$ vary along the  galaxy sequence.  In Appendix~\ref{A1}, we show
that, by using the $R$-band  PHP equation and the $R$-$K$ CM relation,
it is possible to derive an equation which is analogous to that of the
$K$-band  PHP,  except for  a  term  which  is proportional  to  $\log
\frac{n_{\rm  R}}{n_ {\rm  K}}$.  It  turns out,  therefore,  that the
waveband variation  of the  PHP does not  depend significantly  on the
ratio of  effective radii, but it  mainly informs on how  the ratio of
Sersic indices vary with other structural parameters of galaxies.

As  detailed in Appendix~\ref{A1},  assuming that  the ratio  of shape
parameters is only a function of the galaxy luminosity, i.e.
\begin{equation}
\log \frac{n_{\rm R}}{n_{\rm K}} = \omega_1 K + \omega_2, 
\label{EQ1}
\end{equation}
where  $K$ is  the total  luminosity in  the $K$  band, we  obtain two
independent  constraints on  $\omega_1$ from  the $R$-  and  $K$- band
values of $a$ and $b$ respectively. These constraints are
\begin{equation}
\begin{array}{l}
\omega_1 = \frac{a_{\rm R}/a_{\rm K} - 1 - 5 b_{\rm R} \Gamma}{5
  a_{\rm R}},
\end{array}
\label{EQ2a}
\end{equation}
for the coefficient $a$, and
\begin{equation}
\begin{array}{l}
\omega_1 = \frac{(b_{\rm K} - b_{\rm R}) (1+5 b_{\rm R} \Gamma)
}{a_{\rm R} (1 - 5 b_{\rm R})}, 
\end{array}
\label{EQ2b}
\end{equation}
for the coefficient  $b$, where $\Gamma$ is the slope  of the $R-K$ CM
relation\footnote{For    consistency    with    the    procedure    of
Appendix~\ref{A1},  we  calculated  $\Gamma$  by constructing  the  CM
relation with  total galaxy colours,  obtained from the  difference of
Kron magnitudes in the $R$ and  $K$ bands.  This gives $\Gamma = -0.08
\pm  0.02$.  }   of \ms1008.   Using the  values of  $a$ and  $b$ from
Table~\ref{R_PHP} and  Table~\ref{K_PHP}, and taking  into account the
correlation of uncertainties on  PHP coefficients, we obtain $\omega_1
= 0.035 \pm 0.035$  ($\sim 8 \pm 8 \%$) and $\omega_1  = -0.2 \pm 0.2$
($-46 \pm  46 \%$) from  the first and second  equations respectively.
We note  (i) that the  two values of  $\omega_1 $ are  consistent with
each  other, showing that  the ansatz  in Eq.~\ref{EQ1}  is compatible
with the  values of the PHP  slopes, and (ii) that  the first equation
sets  much stronger constrains  on the  value of  $\omega_1$, implying
that  the variation of  $\log \frac{n_{\rm  R}}{n_{\rm K}}$  on galaxy
luminosity is  mainly driven  by the \lsn  \, coefficient of  the PHP.
Since  the value  of  $\omega_1$  is fully  consistent  with zero,  we
conclude  that the  ratio  of optical-NIR  shape  parameters does  not
change or can have only a mild variation with galaxy luminosity.  This
result  can   be  compared  with  the   findings  of  \citet[hereafter
LBM03]{LBM03}, who showed that  the waveband variation of the Kormendy
relation  slope constrains how  the ratio  of effective  radii between
different wavebands vary with  galaxy luminosity.  In other terms, the
waveband dependences  of the  KR slope  and of the  PHP \lsn  \, slope
constrain the variation with luminosity  of the ratios of galaxy radii
and  shape parameters  respectively.  This  is an  interesting result,
since  these ratios fully  characterize the  radial colour  profile in
galaxies.   Since   LBM03  found  that   $\frac{r_{\rm  e,opt}}{r_{\rm
e,NIR}}$  does not  change  significantly with  luminosity, the  above
result  indicates that  all the  properties of  the colour  profile in
galaxies  (that  is  its  shape  and  gradient)  do  not  change  with
luminosity.  The existence of a correlation between the colour profile
and  galaxy  luminosity  is  a  still  debated  issue.   For  example,
\citet{PDI90} found no correlation among the internal colour gradients
of field  ETGs and their luminosities, while  \citet{TaO03} found that
for very bright ETGs in a nearby cluster such correlation could exist.
A steepening of  colour gradients with galaxy luminosity  is a natural
expectation of  the monolithic collapse model  \citep{LAR74} of galaxy
formation, since galactic winds blow earlier in less massive galaxies,
preventing gas  dissipation to carry  heavy metals in the  center, and
producing, therefore, a less steep  gradient in these systems.  On the
other hand,  in a hierarchical  scenario of galaxy  formation, merging
would   dilute  stellar  population   gradients  in   larger  galaxies
\citep{WHI80},    sweeping    out     or    reversing    the    colour
gradient-luminosity  relation.   The   implications  deriving  from  a
multiwavelength analysis  of the  KR and the  PHP should  be carefully
taken  into  account by  any  model  aimed  to explain  the  processes
underlying the formation and evolution of ETGs.

\subsection{The redshift dependence of the PHP}
\label{SD3}
Since the PHP  is a correlation among global  parameters of ETGs which
are  strictly related  both  to their  luminosity  density (\mie)  and
internal structure  (\re \,  and \sn), its  slopes indicate  how these
properties  vary along  the galaxy  sequence.  The  fact that  the PHP
slopes do not  change significantly up to $z\sim  0.3$ implies that in
this redshift range  (i) the luminosity evolution of  (bright) ETGs is
not  significantly  different  along  the  galaxy  sequence  and  (ii)
correlations among structural properties and galaxy mass do not change
significantly  with $z$.   Point  (i) follows  directly  from the  PHP
equation  (Eq.~\ref{PHPEQ}).  If  the evolution  with $z$  of  \mie \,
would depend on  \lre \, and/or \sn, we would expect  to find a change
of $a$ and $b$ with  redshift (see Eq.~\ref{PHPEQ}). The fact that the
luminosity evolution of (bright) ETGs does not change along the galaxy
sequence is  consistent with results  of Kormendy relation  studies at
intermediate  redshifts (e.g.   \citealt{BAS98, ZSB99,  LBM03}), which
found that the  luminosity evolution of ETGs is  almost independent of
galaxy   size.    Point   (ii)   can  be   analyzed   by   considering
Eq.~\ref{SEEQ2}.   If  the  PHP  origins from  a  correlation  between
specific  entropy and galaxy  mass, the  slopes of  the PHP  are fully
characterized by  the slopes  of the  $M/L$ vs.  $M$  and $s$  vs. $M$
relations.  Since FP studies seem  to indicate that the $M/L$ vs.  $M$
relation   does   not   change   significantly   at   $z   \sol   0.3$
(\citealt{KEL00}),  the result  of Section~\ref{PHPZ0}  implies  that the
slope of the $s$ vs.   $M$ relation does not change significantly with
$z$.  This  is consistent with the  idea that merging  is the physical
process which builds up the  $s$--$M$ relation (MLC00) and that bright
ETGs in  clusters are  mostly assembled at  redshift $z \sog  1$ (e.g.
\citealt{KAU95}).

As shown in Section~\ref{PHPZ0}, the zero-point of the PHP can be applied
to constrain the mean luminosity evolution of ETGs with redshift. This
use of the PHP is based on the fact that its slopes do not evolve with
redshift and that  the variations with $z$ of the  mean values of both
\lre  \,  and \lsn  \,  are  known  exactly (see  Eq.~\ref{DELTAC_1}).
Assuming   $\Delta   (\overline{\log   r_{\rm  e}})=0$   and   $\Delta
(\overline{\log n})=0$,  we find that the evolution  of the zero-point
of the  PHP from $z  \sim 0.3$  to $z \sim  0$ is consistent  with the
cosmological dimming of mean surface brightness and the passive fading
of an old  stellar population with a high  formation redshift, $z_{\rm
f} >  1$--$2$.  The hypothesis $\Delta  (\overline{\log r_{\rm e}})=0$
is  well motivated  by  the  findings of  different  studies that  the
distribution  of  radii  of  ETGs  does not  change  significantly  at
intermediate  redshifts  \citep{LBM02,  SMW03}.   On the  other  hand,
assuming $\Delta (\overline{\log n})=0$  is a more tricky point, since
a  morphology-density relation  is known  to exist  and  this relation
evolves with  redshift \citep{DOC97}.  To account for  this effect, we
calculated the mean values of \lsn \, for the sample of \ms1008 \, and
that of GRA02.  The difference between the two values turned out to be
$\Delta (\overline{\log n}) = -0.045 \pm 0.05$\footnote{This value was
obtained by considering  only the galaxies of the  GRA02 sample within
the  same magnitude  range of  the  galaxies in  \ms1008 (i.e.   $\sim
m*+3$).  The  same result,  however, is obtained  by not  applying any
magnitude cut.   In this case  we obtain $\Delta  (\overline{\log n})=
-0.04   \pm  0.06$},   in  agreement   with  the   hypothesis  $\Delta
(\overline{\log n})=0$.

Due  to heavy request  for measuring  velocity dispersions  of distant
galaxies, it is  clearly of great interest to have  shown that the PHP
is a valuable tool for measuring the luminosity evolution of ETGs.  In
the future, it will be very  interesting to derive the PHP for samples
of galaxies at higher redshift, taking advantage (a) of the relatively
small dispersion of  the PHP with respect to  other purely photometric
correlations, and  (b) of  the fact  that the PHP  slopes seem  not to
depend  on  the   waveband,  allowing,  therefore,  a  straightforward
comparison of  the PHP among different  redshifts (different restframe
bands) to be performed.

\section{Summary}
Using  a unique  dataset  of optical  and  NIR data,  we  have done  a
detailed study of the Photometric Plane (PHP) relation in the $R$, $I$
and $K$  bands for a large sample  of ETGs in the  rich galaxy cluster
\ms1008 at intermediate redshift ($z  \sim 0.3$).  We have derived the
PHP by  accounting for selection  effects and other issues  related to
the fit of this relation.
Our main findings can be summarized as follows.

\begin{enumerate}

\item The PHP  is indeed a plane in the space  of the quantities \lre,
\mie  \, and  \lsn, i.e.,  there is  no significant  departure  from a
purely  flat relation  (Section~\ref{OPTPHP}).  We  have shown  that this
result  is  consistent with  the  fact that  the  PHP  origins from  a
systematic  variation of  specific entropy  of ETGs  along  the galaxy
sequence.   For isotropic,  non  rotating, one  component models,  the
specific  entropy turns out,  in fact,  to be  almost an  exact linear
combination of \lre, \mie \, and \lsn.

\item  We  have  found  no  waveband  dependence  of  the  PHP  slopes
(Section~\ref{OPTKKPHP}).   Both the PHP  coefficients and  its intrinsic
scatter are fully consistent among the optical and NIR wavebands.

\item  The scatter  around the  PHP is  about $\sim  32$ \%,  which is
larger than the average scatter  of the Fundamental Plane, but smaller
than that  of the Kormendy relation  and the photometric  or the `$\rm
Mg_2$ Fundamental Plane' (Section~7.1).

\vspace{0.2cm} By comparing the PHP coefficients at $z=0.3$ with those
at $z \sim 0$  for the sample of GRA02, we have  found that the slopes
of the PHP do not show a significant variation in this redshift range.
This  fact has  important  consequences on  our  understanding of  the
evolution of bright ETGs, as discussed in Section~7.4.  Namely:
\item the luminosity evolution does not change significantly along the
galaxy sequence;
\item the  slope of the  relation between specific entropy  and galaxy
  mass does not change since $z \sim 0.3$.

Both these  findings are in agreement  with the fact  that bright ETGs
are already assembled at high redshift.  Finally, we have demonstrated
that the PHP is a valuable tool for measuring the luminosity evolution
of ETGs. We have found that
\item  the mean  luminosity of  ETGs  is consistent  with the  passive
fading of  an old stellar  population, with formation redshift  $z_{\rm f} >
1$.

\end{enumerate}

 Due to  the heavy  request of telescope  time for  measuring velocity
dispersions,  constructing  the  FP  relation  for  large  samples  of
galaxies at $z > 0.5$ becomes impracticable.  The present results show
that  the PHP could  be a  very interesting  alternative tool  for the
study of ETGs at high redshifts.

%
%

\section*{Acknowledgments}
We thank R.R.   de Carvalho for the careful  reading of our manuscript
and for the helpful suggestions.   We also thank the anonymous referee
for the helpful comments and suggestions.  FLB thanks the INAF for the
post-doc fellowship.   GC is supported by the  RTN Euro3D postdoctoral
fellowship,  funded  by the  European  Commission  under contract  No.
HPRN-CT-2002-003005.   CPH and AM  acknowledge the  financial supports
provided  through the  European Community's  Human  Potential Program,
under  contract  HPRN-CT-2002-0031  SISCO,  and the  Regione  Campania
(L.R. 05/02)  project {\it `Evolution of Normal  and Active Galaxies'}
respectively.  This  work has been partially supported  by the Italian
Ministry   of  Education,  University,   and  Research   (MIUR)  grant
COFIN2003020150:  {\it  Evolution of  Galaxies  and Cosmic  Structures
after the Dark Age: Observational Study}.


\appendix

\section{PSF modeling}
\label{PSF}
\begin{figure*}
\centering
\includegraphics[angle=0,width=15cm,height=15cm]{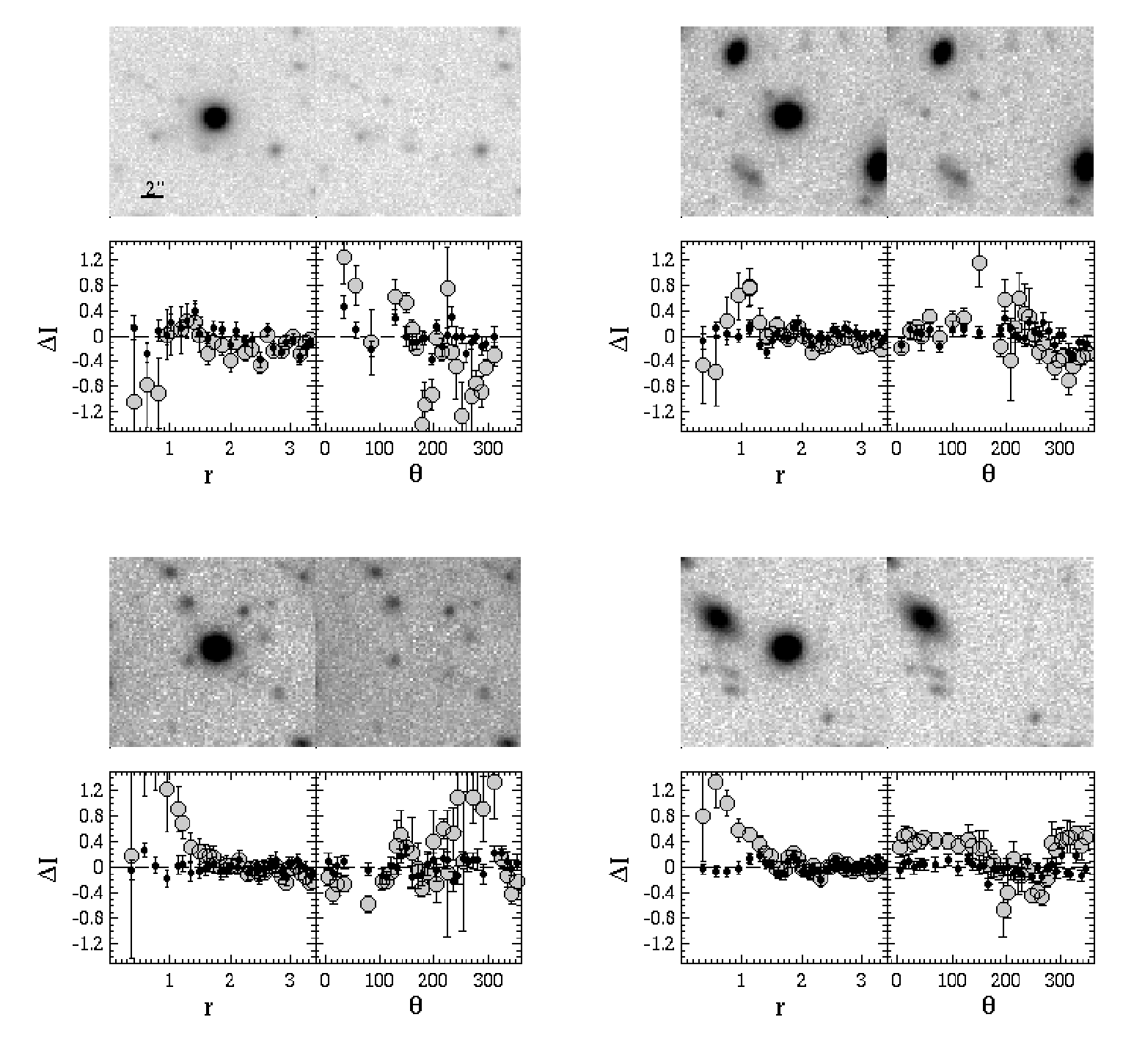}
\caption{PSF fitting to four stars  in the FORS1 $R$-band image.  Each
plot shows the image of  the star (upper-left panel), the residual map
(upper-right), and  mean fitting residuals in units  of the background
standard deviation as a function of the radial and angular coordinates
(lower-left and lower-right  panels, respectively). Large gray symbols
and small black circles  denote the residuals obtained by respectively
excluding and including the angular modulation function in the PSF fit
(see text for  details).  The spatial scale of each  image is the same
of that shown in the upper--left plot. }
\label{PSFFIT}
\end{figure*}

The stars in the $R$, $I$, \KS \, and \KI \, images were fitted by the
following formula:
\begin{equation}
\mathrm{ PSF(r) = \sum_{k=1}^p A_k(\theta) \cdot M_k(r, \theta),
\label{PSFMOD} }
\end{equation}
where $r$  is the distance to  the star center, $\theta$  is the polar
angle, $M  \propto \left[ 1+(r/r_{\rm c})^2 \right]^{-  \beta}$ is the
Moffat law~\citep{MOF69},  $p$ is the number of  Moffat functions, and
$A$ is an angular modulation function.  PSF asymmetries were described
by adopting the formula:
\begin{equation}
  A(\theta)  = \sum_{m=1}^{q} \left[  a_m \cos(m  \cdot \theta)  + b_m
 \sin(m \cdot \theta) \right].
\end{equation}
The values of $p$ and $q$  were chosen interactively for each image in
order to obtain  an accurate modeling of stellar  isophotes.  This was
achieved  by using  two or  three Moffat  functions, depending  on the
image,  with $q  \le 3$.   To account  for PSF  variations  across the
frames, each  galaxy was fitted by  using a local  PSF model, obtained
from the nearest star in the  field.  The PSF modeling in the $R$ band
image is illustrated in  Fig.~\ref{PSFFIT}, where we show residuals of
star fitting as a function of both the distance to the star center and
the polar  angle.  Residuals are  plotted for two fitting  cases, with
and without  the angular modulation  function.  The figure  shows that
the adopted models  give an accurate description of  the PSF, allowing
both  the  radial  and  the   angular  behaviour  of  the  PSF  to  be
reproduced. We note that the use of asymmetric terms in the fit allows
a  better  centering of  the  PSF model  to  be  obtained.  Hence  the
improvement in the radial trend  of PSF residuals which is observed in
Fig.~\ref{PSFFIT}.  Similar results were also obtained in the analysis
of the images of \ms1008 \, in the other bands.

\section{Uncertainties on \re, \mie \, and \sn}
\label{UNCER}
Uncertainties on structural parameters  were estimated by applying the
2D fitting method  to simulated galaxy images. For  each galaxy in the
$R$, $I$, \KS \, and \KI  \, bands, we used the best-fitting values of
\re, \mie \, and \sn \, to construct a set of seeing--convolved Sersic
models to which  both photon and read-out noise  were added.  In order
to account for uncertainties due to PSF modeling and to image masking,
each simulation  was obtained by  using a different PSF,  according to
the uncertainties on the PSF  fitting parameters, while the 2D fitting
was performed  by using  the same PSF  and the  same mask of  the real
galaxies.   For each  galaxy, uncertainties  on  structural parameters
were  estimated   from  the  covariance  matrix   of  2D  best-fitting
parameters  for  the  corresponding  set  of  simulations.   The  mean
uncertainties  on  \lre,  \mie \,  and  \lsn  \,  amount to  $\sim  \!
0.03$~dex,   $\sim  \!   0.18$~$\rm   mag/arcsec^{2}$  and   $\sim  \!
0.05$~dex  respectively, in  $R$;  to $\sim  \!   0.05$~dex, $\sim  \!
0.26$~$\rm mag/arcsec^{2}$ and $\sim \!  0.09$~dex in $I$; to $\sim \!
0.06$~dex,   $\sim  \!   0.27$~$\rm   mag/arcsec^{2}$  and   $\sim  \!
0.06$~dex in  \KI; and  to $\sim \!   0.07$~dex, $\sim  \!  0.34$~$\rm
mag/arcsec^{2}$ and $\sim 0.08$~dex in \KS.

\begin{figure*}
\centering
\includegraphics[angle=0,width=15cm,height=6.0cm]{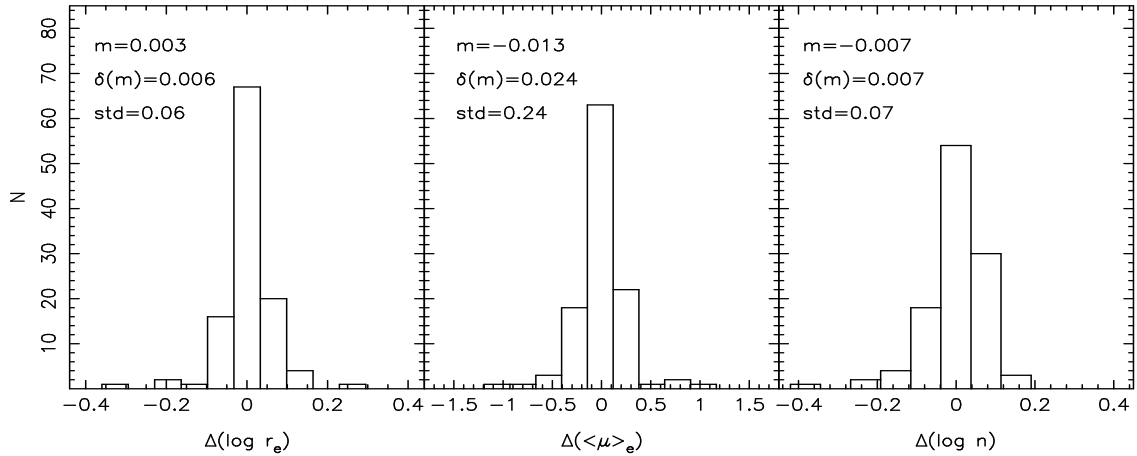}
\caption{Comparison  of  structural  parameters  for the  $\rm  N=112$
  galaxies  in common  between  the $R$-  and  $I$-band samples.   The
  quantities  $m$,  $\delta(m)$ and  $std$  are  the  mean value,  the
  uncertainty  on  the  mean,  and  the standard  deviation  for  each
  distribution. }
\label{confRI}
\end{figure*}

\begin{figure*}
\centering
\includegraphics[angle=0,width=15cm,height=6.0cm]{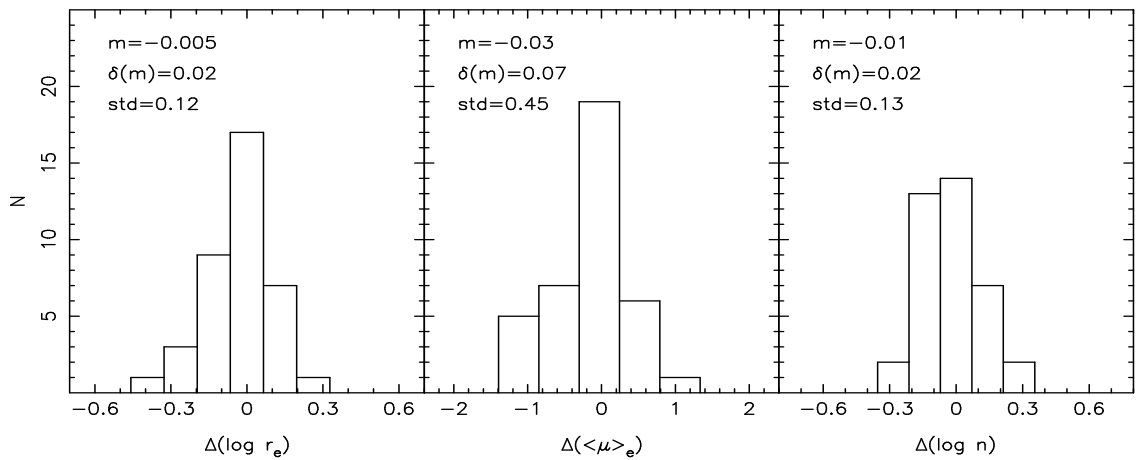}
\caption{The  same of  Fig.~\ref{confRI}  for the  $N=38$ galaxies  in
  common between the \KS \, and \KI \, samples.}
\label{confKK}
\end{figure*}

In  Fig.~\ref{confRI} we compare  the structural  parameters of  the $
N=112$ galaxies in common between the  $R$ and $I$ bands. We note that
the mean  differences of the  distributions are fully  consistent with
zero, in agreement with the fact that  $R$ and $I$ bands at $z \sim \!
0.3$  sample  a very  similar  spectral  region  and ETGs  have  small
optical--optical colour gradients.  For this reason, the widths of the
distributions  in Fig.~\ref{confRI}  provide a  rough estimate  of the
mean uncertainty  on \lre,  \mie \, and  \lsn. These values  are fully
consistent with those obtained by  adding in quadrature the above mean
uncertainties  on  structural  parameters.   In  Fig.~\ref{confKK}  we
compare the  structural parameters of  the $ N=38$ galaxies  in common
between the \KS \, and the \KI \, samples. Again, the mean differences
of the distributions  are fully consistent with zero,  showing that no
significant  systematic effects are  present.  We  also note  that the
standard  deviations  shown   in  Fig.~\ref{confKK}  agree  with  what
expected on the basis of the mean errors on structural parameters (see
above).

\section{Specific entropy of ETGs}
\label{AENT}
As shown by MLC00, the specific  entropy of a galaxy, $s$, is given by
the following formula:
\begin{equation}
s = M_{\rm T}^{-1} \int \left( \ln P^{3/2} \rho^{-5/2} \right) \rho \,
d V,
\label{SEEQ}
\end{equation}
where  $\rho$  and  $P$  are  the  3D  density  and  pressure  fields,
respectively, while $  M_{\rm T}$ is the total  mass.  Following LGM99
and  MLC00, we  consider spherical,  non-rotating galaxy  models, with
negligible  radial  gradients of  the  mass-to-light  ratio. For  such
models the function $\rho(r)$ is  obtained directly from the 2D Sersic
law by solving the Abel integral equation, while $P(r)$ is calculated
from the equation  of hydrostatic equilibrium, $\frac{ d  \!  P }{d \!
r }= - G M \rho/ r^2$,  where the mass profile $M(r)$ is obtained from
a direct  integration of  $\rho(r)$.  The functions  $\rho(r)$, $M(r)$
and  $P(r)$ can  be written  as  products of  dimensional factors  for
dimensionless functions:
\begin{eqnarray}
\rho(r) = M/L  \cdot I_0 / R_{\rm e} \cdot \tilde{\rho}(s) \\
M(r) = M/L \cdot I_0 \cdot R_{\rm e}^2 \cdot \tilde{M}(s) \\
P(r) = G (M/L  \cdot I_0)^2  \cdot \tilde{P}(s)
\end{eqnarray}
where $s=r/R_{\rm e}$ is the  dimensionless radius and $I_0$ is the 2D
central surface brightness.  By  substituting the previous formulae in
Eq.~\ref{SEEQ}, and by using the relation  between $I_0$ and $ < \!  I
\!    >_{\rm   e}$   (Eq.~\ref{magn}),  we   obtain   Eqs.~\ref{SEEQ2}
and~\ref{SEEQ3}.

\section{Optical-NIR PHP}
\label{A1}
Considering the  optical-NIR colour-magnitude relation,  $R-K = \Gamma
\cdot K  + \Delta$, where $R-K$  and $K$ are total  galaxy colours and
magnitudes,   and   using    the   definition   of   total   magnitude
(Eq.~\ref{magn}), we  can write the R-band PHP  relation, $\log R_{\rm
e,R} = a_{\rm  R} \cdot \log n_{\rm R}  + b_{\rm R} \cdot <  \! \mu \!
>_{\rm  e,R} + c_{\rm  R}$, where  the subscript  $\rm R$  denotes the
waveband, as follows:
\begin{equation}
\begin{array}{l}
\log  \frac{R_{\rm e,R}}{R_{\rm e,K}}   +  \log   R_{\rm e,K}  =  \\
a_{\rm R}  \log
\frac{n_{\rm R}}{n_{\rm K}}+a_{\rm R} \log n_{\rm K}  + b_{\rm R}
\left( < \! \mu  \!  >_{\rm e,R} - <
\!  \mu \!  >_{\rm e,K} \right) + \\ +  b_{\rm R} < \! \mu \!  >_{\rm
  e,K} +c_{\rm R} = \\
= a_{\rm R} \log \frac{n_{\rm R}}{n_{\rm K}} + a_{\rm R} \log n_{\rm
  K} + b_{\rm R} \left[ \Gamma < \!
\mu \! >_{\rm e,K} - 5 \Gamma \log R_{\rm e,K} \right. + \\ \left. -2.5 \Gamma
\log (2 \pi) + \Delta + 5 \log \frac{R_{\rm e,R}}{R_{\rm e,K}} \right]
+ b_{\rm R} <
\! \mu \! >_{\rm e,K} + c_{\rm R}.  \\
\end{array}
\end{equation}
From this equation, we obtain:
\begin{equation}
\begin{array}{l}
\log R_{\rm e,K} = \left( 1+ 5 b_{\rm R} \Gamma \right)^{-1} \cdot
\left\{ a_{\rm R} \log \frac{n_{\rm R}}{n_{\rm K}} + (5 b_{\rm R} - 1)
\log \frac{R_{\rm e,R}}{R_{\rm e,K}} + \right. \\
\left. a_{\rm R}   \log n_{\rm K} + b_{\rm R} (1 + \Gamma) < \! \mu \!
>_{\rm e,K} + \left[ c_{\rm R} +b_{\rm R} \Delta -2.5 \Gamma \log (2 \pi) \right] \right\}.
\end{array}
\label{WEQ}
\end{equation}
We note that since $(5 b_{\rm R} - 1) \sim 0$ (see Table~\ref{R_PHP}),
the term  $\log \frac{R_{\rm e,R}}{R_{\rm e,K}}$  in Eq.~\ref{WEQ} can
be  neglected.   Moreover,  since  $\left(  1+  5   b_{\rm  R}  \Gamma
\right)^{-1} \cdot (1  + \Gamma) \sim 1$, we  obtain from the previous
equation:
\begin{equation}
\begin{array}{lll}
\log R_{\rm e,K} & \simeq & \left( 1 + 5 b_{\rm R} \Gamma \right)^{-1}
 \cdot a_{\rm R}
 \log \frac{n_{\rm R}}{n_{\rm K}} + \left( 1 + 5 b_{\rm R} \Gamma \right)^{-1} \cdot \\
    & & \cdot a_{\rm R}
\log n_{\rm K} + b_{\rm R} < \! \mu \! >_{\rm e,K} + const.
\end{array}
\label{RK_EQ}
\end{equation}
We note that  this equation is similar  to that of the PHP  in the $K$
band, except  for a term which  is proportional to the  ratio of shape
parameters between the $R$ and  $K$ bands.  Substituting the ansatz of
Eq.~\ref{EQ1} in the previous equation, we obtain a relation identical
to that  of NIR PHP.  Comparing  the slopes of this  relation with the
values  of $a$  and $b$  in the  $K$ band,  we  obtain Eqs.~\ref{EQ2a}
and~\ref{EQ2b}, respectively.

\begin{figure*}
\centering
\includegraphics[angle=0,width=19cm,height=24cm]{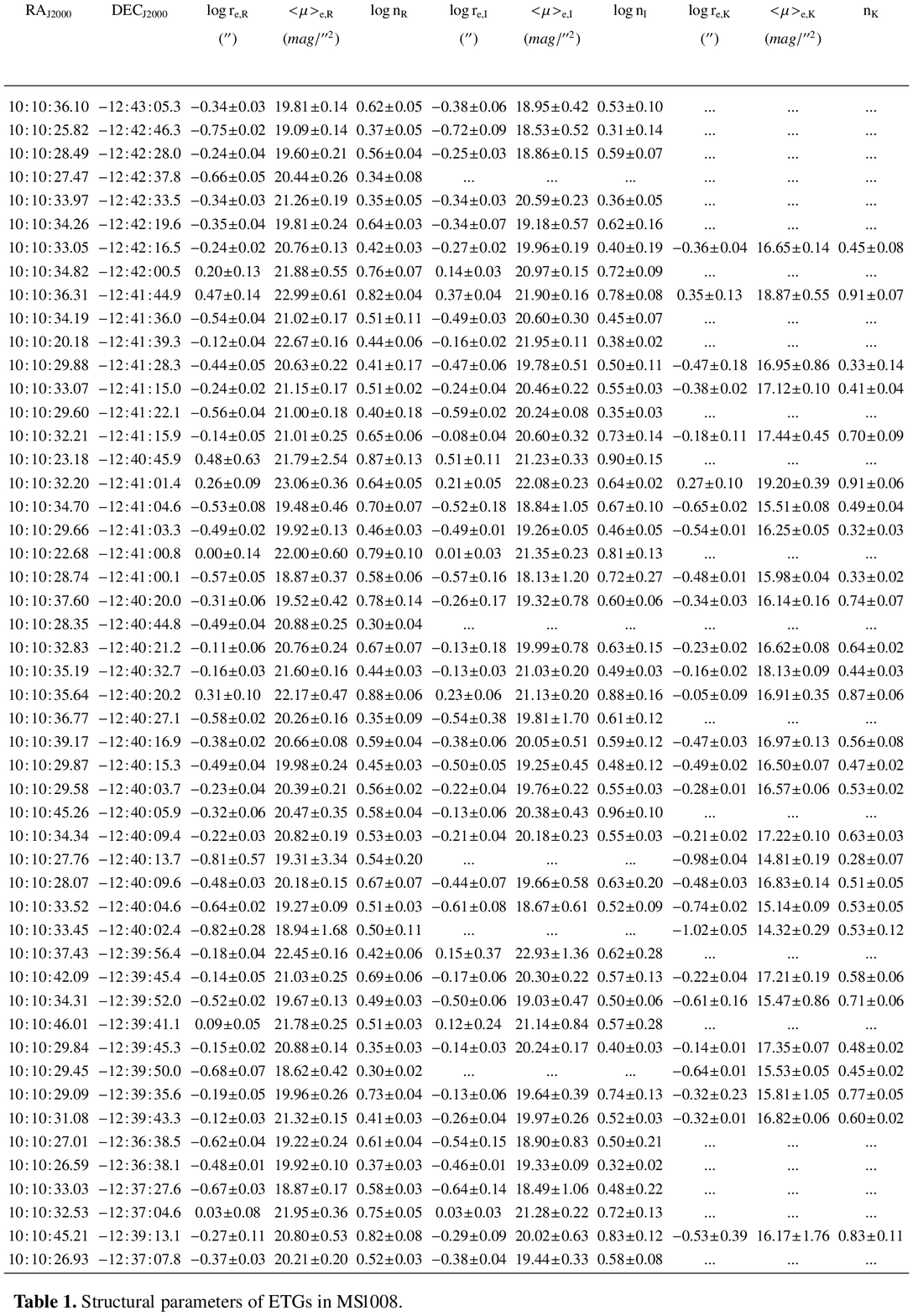}
\end{figure*}
\begin{figure*}
\centering
\includegraphics[angle=0,width=19cm,height=24cm]{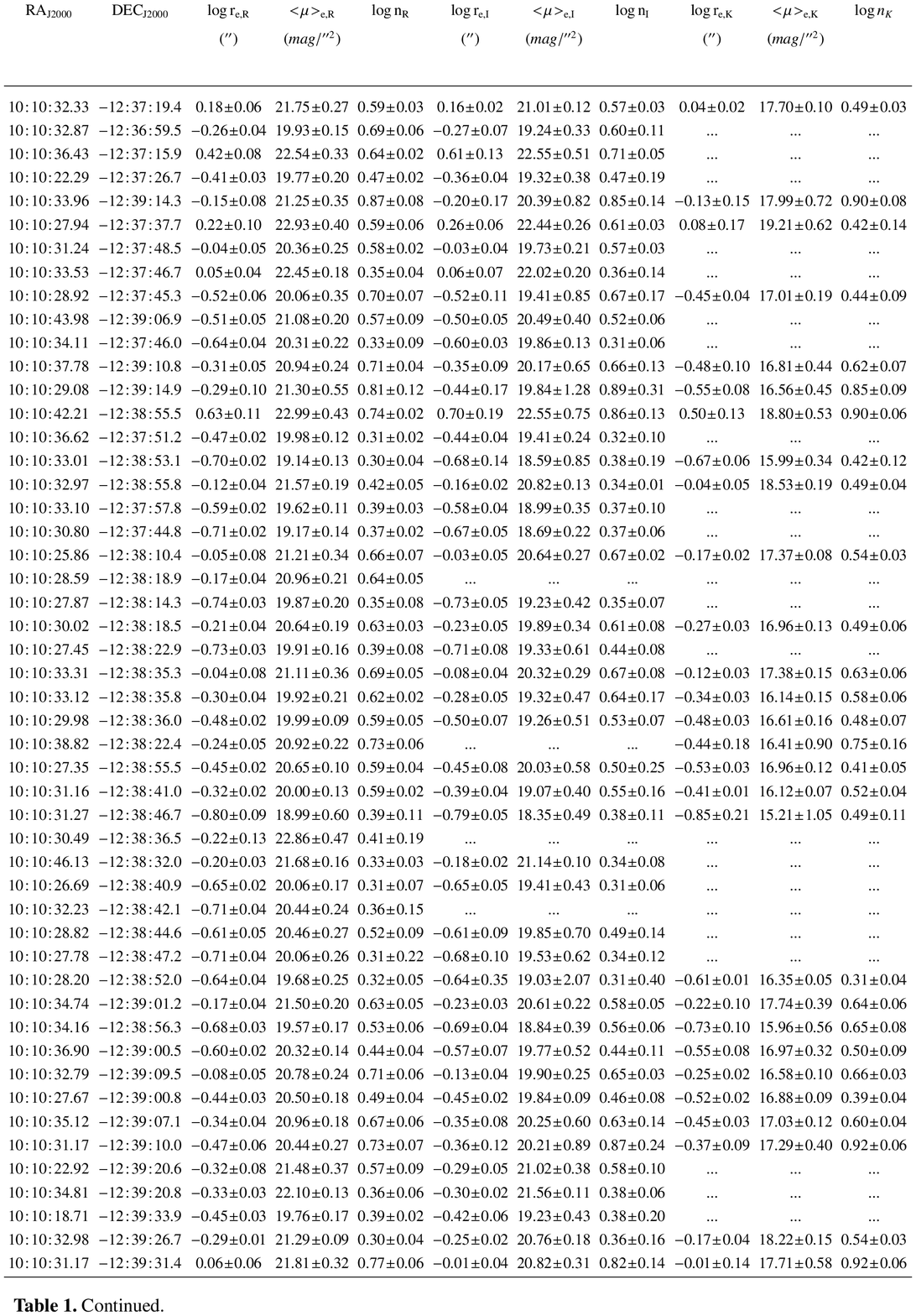}
\end{figure*}
\begin{figure*}
\centering
\includegraphics[angle=0,width=19cm,height=24cm]{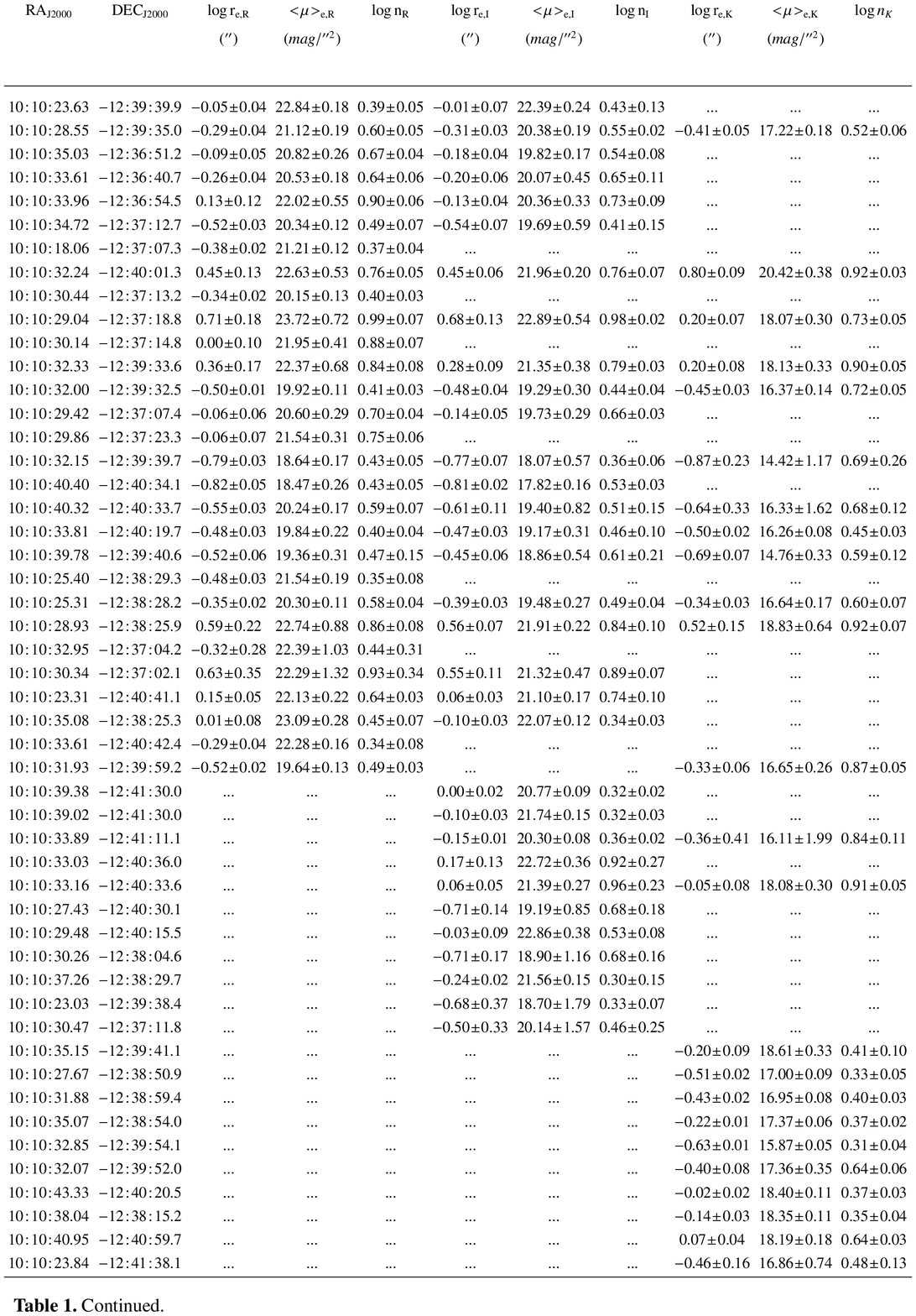}
\end{figure*}



%
%
%
%
%
%

\bsp

\label{lastpage}


\begin{thebibliography}{99}
\bibitem[\protect\citeauthoryear{Barger  et al.}{1998}]{BAS98} Barger,
A.J.,  Arag\'on-Salamanca A.,  Smail I.,  Ellis R.S.,  Couch W.J.,
Dressler A., Oemler A.,  Poggianti B.M., Sharples R.M., 1998, ApJ,
501, 522
\bibitem[\protect\citeauthoryear{Bruzual   \&   Charlot}{2003}]{BrC03}
  Bruzual G., Charlot S., 2003, MNRAS, 344, 1000
\bibitem[\protect\citeauthoryear{Bertin    \&    Arnout}{1996}]{BeA96}
Bertin E., Arnout S., 1996, A\&AS, 117, 393
\bibitem[\protect\citeauthoryear{Busarello    et    al.}{2002}]{BML02}
  Busarello G., Merluzzi P.,  La Barbera F., Massarotti M., Capaccioli
  M., 2002, A\&A, 389, 787
\bibitem[\protect\citeauthoryear{Busarello    et    al.}{1997}]{BUS97}
  Busarello  G., Capaccioli M.,  Capozziello S.,  Longo G.,  Puddu E.,
  1997, A\&A, 320, 415
\bibitem[\protect\citeauthoryear{Caon,          Capaccioli          \&
    D'Onofrio}{1993}]{CCD93}  Caon N.,  Capaccioli  M., D'Onofrio  M.,
    1993, MNRAS, 265, 1013
\bibitem[\protect\citeauthoryear{Capaccioli}{1989}]{CAP89}   Capaccioli
M., 1989, in:  The world of galaxies, eds.  H.G. Corwin, L. Bottinelli
Springer-Verlag, Berlin, p. 208
\bibitem[\protect\citeauthoryear{Ciotti    \&    Bertin}{1999}]{CIO99}
  Ciotti L., Bertin G., 1999, A\&A, 352, 447
\bibitem[\protect\citeauthoryear{de   Propris   et   al.}{1999}]{dP99}
de Propris R., Stanford S.A., Eisenhardt P.R., Dickinson M., Elston
R., 1999, AJ, 118, 719
\bibitem[\protect\citeauthoryear{de             Carvalho            \&
Djorgovski}{1989}]{dCD89}  de Carvalho  R.R.,  Djorgovski S.,  1989,
ApJ, 341, 37
\bibitem[\protect\citeauthoryear{Djorgovski \& Davies}{1987}]{DjD87}
Djorgovski S., Davies M., 1987, ApJ, 312, 59
\bibitem[\protect\citeauthoryear{Dressler     et    al.}{1987}]{DLB87}
Dressler A., Lynden-Bell D., Burstein D., Davies R.L., Faber S.M.,
Terlevich R., Wegner G., 1987, ApJ, 313, 42
\bibitem[\protect\citeauthoryear{Dressler     et    al.}{1997}]{DOC97}
Dressler,  A., Oemler,  A.Jr., Couch,  W.J., Smail,  I.,  Ellis, R.S.,
Barger, A.,  Butcher, H., Poggianti, B.M., Sharples,  R.M., 1997, ApJ,
490, 577
\bibitem[\protect\citeauthoryear{Gioia \& Luppino}{1994}]{GiL94} Gioia
I.M., Luppino G.A., 1994, ApJS, 94, 583
\bibitem[\protect\citeauthoryear{Gioia   et  al.}{1990}]{GMS90}  Gioia
I.M., Maccacaro T.,  Schild R.E., Wolter A., Stocke  J.T., 1990, ApJS,
72, 567
\bibitem[\protect\citeauthoryear{Graham}{2002}]{GRA02}  Graham A.W., 2002,
MNRAS, 334, 859 (GRA02)
\bibitem[\protect\citeauthoryear{J\o    rgensen,   Franx    \&   Kj\ae
rgaard}{1996}]{JFK96} J\o rgensen I., Franx M., Kj\ae rgaard P., 1996,
MNRAS, 2809, 167
\bibitem[\protect\citeauthoryear{Kauffmann}{1995}]{KAU95}     Kauffmann
G., 1995, MNRAS, 274, 161
\bibitem[\protect\citeauthoryear{Kelson  et al.}{2000}]{KEL00} Kelson
D.D., Illingworth G.D., van Dokkum  P.G., Franx M., 2000, ApJ, 531,
184
\bibitem[\protect\citeauthoryear{Khosroshahi   et   al.}{2000}]{KWK00}
Khosroshahi  H., Wadadekar Y.,  Kembhavi A.,  Mobasher B.,  2000, ApJ,
531, L103
\bibitem[\protect\citeauthoryear{Khosroshahi   et   al.}{2004}]{KRP04}
Khosroshahi H.,  Raychaudhury S.,  Ponman T. J.,  Miles T.  A., Forbes
D. A.  2004, MNRAS, 349, 527
\bibitem[\protect\citeauthoryear{Kormendy}{1977}]{KOR77}  Kormendy J.,
1977, ApJ, 218, 333
\bibitem[\protect\citeauthoryear{La         Barbera,        Busarello,
Capaccioli}{2000}]{LBM00} La Barbera  F., Busarello G., Capaccioli M.,
2000, A\&A, 362, 851 (LBC00)
\bibitem[\protect\citeauthoryear{La Barbera et al.}{2002}]{LBM02}  La
Barbera F., Busarello G., Merluzzi P., Massarotti M., 2002,
ApJ, 571, 790
\bibitem[\protect\citeauthoryear{La  Barbera et  al.}{2003}]{LBM03} La
Barbera F., Busarello G., Merluzzi  P., Massarotti M., 2003, ApJ, 595,
127 (LBM03)
\bibitem[\protect\citeauthoryear{La  Barbera et  al.}{2004}]{LMB04} La
Barbera  F., Merluzzi P.,  Busarello G.,  Massarotti M.,  Mercurio A.,
2004, A\&A, 425, 797 (LMB04)
\bibitem[\protect\citeauthoryear{Larson}{1974}]{LAR74}  Larson  R.B.,
1974, MNRAS, 166, 585
\bibitem[\protect\citeauthoryear{Lewis   et  al.}{1999}]{LEW99}  Lewis
A.D., Ellingson E., Morris S.L., Carlberg R.G., 1999, ApJ, 517, 587
\bibitem[\protect\citeauthoryear{Lima Neto  et al.}{1999}]{LGM99} Lima
Neto G.B., Gerbal D., M\' arquez I., 1999, MNRAS, 309, 481 (LGM99)
\bibitem[\protect\citeauthoryear{M\'arquez    et    al.}{2000}]{MLC00}
M\'arquez L., Lima Neto G.B., Capelato H., Durret E., Gerbal D., 2000,
A\&A, 353, 873 (MLC00)
\bibitem[\protect\citeauthoryear{M\'arquez    et    al.}{2001}]{MLC01}
M\'arquez L.,  Lima Neto  G.B., Capelato H.,  Durret E.,  Lanzoni, B.,
Gerbal D., 2001, A\&A, 379, 767 (MLC01)
\bibitem[\protect\citeauthoryear{Moffat}{1969}]{MOF69} Moffat A.F.J., 
1969, A\&A, 3, 455
\bibitem[\protect\citeauthoryear{Pahre,      Djorgovski      \&     de
Carvalho}{1998}]{PDdC98}  Pahre  M.A.,  Djorgovski S.G.,  de  Carvalho
R.R., 1998, AJ, 116, 1606
\bibitem[\protect\citeauthoryear{Peletier     et    al.}{1990}]{PDI90}
  Peletier  R.F.,  Davies  R.L.,  Illingworth  G.D.,  Davis  L.E.,
  Cawson M., 1990, AJ, 100, 1091
\bibitem[\protect\citeauthoryear{Prugniel   \&   Simien}{1996}]{PrS96}
Prugniel P., Simien F., 1996, A\&A, 309, 749
\bibitem[\protect\citeauthoryear{Scalo}{1986}]{SCALO}   Scalo   M.J.,
1986, Fundamentals of Cosmic Physics, 11, 1
\bibitem[\protect\citeauthoryear{Sersic}{1968}]{SER68}          Sersic
J.L.,  1968,  Atlas de  Galaxias  Australes, Observatorio  Astronomico,
Cordoba
\bibitem[\protect\citeauthoryear{Shen et  al.}{2003}]{SMW03} Shen S.,
Mo  H.J., White  S.D.M., Blanton  M.R., Kauffmann  G.,  Voges W.,
Brinkmann J., Csabai I., 2003, MNRAS, 343, 978
\bibitem[\protect\citeauthoryear{Tamura \& Ohta}{2003}]{TaO03} Tamura
    N., Ohta K. 2003, AJ, 126, 596
\bibitem[\protect\citeauthoryear{Tran   et   al.}{2004}]{TRAN04}  Tran
K.H., Franx M., Illingworth G.D.,  van Dokkum P.G., Kelson D.D., Magee
D., 2004, ApJ, 609, 683
\bibitem[\protect\citeauthoryear{van Dokkum  et al.}{1998}]{vDF98} van
Dokkum, P.G., Franx, M.,  Kelson, D.D., Illingworth, G.D., Fisher, D.,
and Fabricant, D. 1998, ApJ, 500, 714
\bibitem[\protect\citeauthoryear{Yee    et   al.}{1998}]{YEM98}   Yee
  H.K.C., Ellingson E., Morris  S.L., Abraham R.G., Carlberg R.G.,
  1998, ApJS, 116, 211
\bibitem[\protect\citeauthoryear{White}{1980}]{WHI80}   White   S.D.,
1980, MNRAS, 191, 1
\bibitem[\protect\citeauthoryear{Wuyts et al.}{2004}]{WUY04} Wuyts S.,
van Dokkum P.G., Kelson D.D., Franx M., Illingworth G.D., 2004, ApJ,
605, 677
\bibitem[\protect\citeauthoryear{Ziegler     et     al.}{1999}]{ZSB99}
Ziegler B.,  Saglia R., Bender  R., Belloni P., Greggio  L., Seitz
S., 1999, A\&A, 346, 13
\end{thebibliography}
\end{document}